\documentclass[10pt,journal,compsoc]{IEEEtran}
\ifCLASSOPTIONcompsoc
  \usepackage[nocompress]{cite}
\else
  \usepackage{cite}
\fi

\ifCLASSINFOpdf
\else
\fi

\usepackage[cmex10]{amsmath}
\usepackage{array}
\usepackage{amsmath}
\usepackage{mdwmath}
\usepackage{mdwtab}
\usepackage{eqparbox}
\usepackage{url}
\usepackage{xcolor,soul,framed} 
\usepackage{graphicx}
\usepackage{pifont}
\usepackage{makecell}
\usepackage{multirow}
\usepackage{amssymb}
\usepackage{algorithmic}
\usepackage[linesnumbered,longend,ruled,vlined]{algorithm2e}

\begin{document}
%
\title{FIRED: a fine-grained robust performance diagnosis framework for cloud applications}
%
%
%
%

\author{Ruyue Xin,~\IEEEmembership{Student Member,~IEEE,}
        Hongyun Liu,~\IEEEmembership{Student Member,~IEEE,}
        Peng Chen,~\IEEEmembership{Member, ~IEEE,}
        Paola Grosso,~\IEEEmembership{Member,~IEEE}
        Zhiming Zhao,~\IEEEmembership{Senior Member,~IEEE}
\IEEEcompsocitemizethanks{\IEEEcompsocthanksitem The authors are with Multiscale Networked Systems (MNS), University of Amsterdam, the Netherlands.\protect
\IEEEcompsocthanksitem Corresponding authors: Peng Chen (chenpeng@mail.xhu.edu.cn) and Zhiming Zhao (z.zhao@uva.nl).
\IEEEcompsocthanksitem Peng Chen is also with School of Computer and Software Engineering, Xihua University, Chengdu, China.
}
\thanks{Manuscript received ; .}}

%
%

\markboth{Journal of \LaTeX\ Class Files,~Vol.~14, No.~8, August~2015}%
{Shell \MakeLowercase{\textit{et al.}}: Bare Demo of IEEEtran.cls for Computer Society Journals}
%



\IEEEtitleabstractindextext{%
\begin{abstract}
To run a cloud application with the required service quality, operators have to continuously monitor the cloud application's run-time status, detect potential performance anomalies, and diagnose the root causes of anomalies. However, existing models of performance anomaly detection often suffer from low re-usability and robustness due to the diversity of system-level metrics being monitored and the lack of high-quality labeled monitoring data for anomalies. Moreover, the current coarse-grained analysis models make it difficult to locate system-level root causes of the application performance anomalies for effective adaptation decisions. We provide a FIne-grained Robust pErformance Diagnosis (FIRED) framework to tackle those challenges. The framework offers an ensemble of several well-selected base models for anomaly detection using a deep neural network, which adopts weakly-supervised learning considering fewer labels exist in reality. The framework also employs a real-time fine-grained analysis model to locate dependent system metrics of the anomaly. Our experiments show that the framework can achieve the best detection accuracy and algorithm robustness, and it can predict anomalies in four minutes with F1 score higher than 0.8. In addition, the framework can accurately localize the first root causes, and with an average accuracy higher than 0.7 of locating first four root causes.
\end{abstract}

\begin{IEEEkeywords}
Performance diagnosis, weakly-supervised, deep ensemble learning, dependency graph, random walk.
\end{IEEEkeywords}}

\maketitle

\IEEEdisplaynontitleabstractindextext

%
\IEEEpeerreviewmaketitle

\IEEEraisesectionheading{\section{Introduction}\label{sec:introduction}}
\IEEEPARstart{C}{loud} environments provide elastic and on-demand resources for developing applications\cite{huan2019}. However, because of the inherent dynamism of clouds, performance anomalies of cloud applications such as degraded response time caused by resource saturation may severely affect the quality of the user experience. In addition, considering complex dependencies and multiple components in cloud applications, it's difficult for operators to detect performance anomalies and identify root causes. Traditionally, operators perform diagnoses for cloud applications manually, which is complicated and time-consuming. Data of different monitoring metrics, e.g., CPU and memory usage, can be continuously collected, reflecting the run-time status of cloud applications \cite{zhao2020multivariate}. Therefore, we could consider a performance diagnosis solution that leverages monitoring data and supports rapid recovery and loss mitigation of cloud applications.

For a real cloud application, monitoring data can be identified as application and system-level data. Application-level data such as response time can be used to detect performance anomalies when slow response time is defined as an anomaly. However, it's hard to capture the status of underlying cloud environments and exploit root causes of performance anomalies with single-variate application-level data. Meanwhile, underlying resources affect the performance of cloud applications heavily. System-level data mainly includes underlying resources, such as CPU, memory, disk, and network. A single resource metric may not precisely reflect the health status of the whole application. Therefore, it's reasonable to detect performance anomalies in cloud applications based on all system-level monitoring information, which is multi-variate time series data. In addition, when a performance anomaly occurs, we need to fine-grained pinpoint root causes on system-level data, which is helpful for the rapid recovery of a cloud application. 

Performance diagnosis is a process of detecting abnormal performance phenomena, e.g., degradation, predicting anomalies to forestall future incidents, and localizing the causes of performance anomalies \cite{ibidunmoye2015performance}. In recent years, studies for performance diagnosis have been developed and mainly focus on performance anomaly detection and root cause localization. For performance anomaly detection, numerous existing methods \cite{hu2018detecting}\cite{breunig2000lof} target improving detection accuracy, but their performance is inconsistent in cloud environments. For example, scaling of cloud infrastructures will change the data distributions of monitoring data, severely affecting detection performance. Therefore, robust performance anomaly detection is necessary for performance diagnosis to keep performance consistency. As for root cause localization, approaches are still developing \cite{gan2019seer}\cite{brandon2020graph} and most of them are coarse-grained, which focuses on service-level or container-level faulty \cite{wu2020microrca}\cite{yu2021microrank}. To fill these gaps, we are motivated to develop a performance diagnosis, which can detect performance anomalies with good robustness and identify the root causes with fine-grained.


As for the performance diagnosis framework, we identify several challenging requirements for performance anomaly detection and root cause localization based on real scenarios. 
\begin{itemize}
    \item Anomalies need to be detected accurately. The detection should also have good robustness to perform consistently for different patterns in monitoring data.
    \item Multi-step prediction of future anomalies is necessary to effectively prevent potential application violations.
    \item When an anomaly occurs, root causes need to be localized in real-time with fine-grained, which is more helpful for efficient application maintenance.
\end{itemize}
Furthermore, the development of the diagnosis framework has to handle two data challenges: 
\begin{itemize}
    \item Monitoring data usually contains only fewer labels or no labels that can be immediately used to train a machine learning-based model because labeling data is often manual and time-consuming.
    \item Collected monitoring data of cloud applications usually contains noise, which can influence the performance of anomaly detection and root cause localization.
\end{itemize}

To address the two data challenges, we adopt weakly-supervised learning \cite{zhou2018brief} and provide methods to filter data noise. In addition, to satisfy the three requirements, we consider the performance diagnosis framework should have good detection performance, multi-step anomaly prediction ability, and fine-grained root cause localization. As for the performance anomaly detection and prediction, we consider integrating existing detection methods based on ensemble learning \cite{galicia2019multi} to improve the detection performance. As for root cause localization, causality inference and graph methods \cite{arya2021evaluation} can be used. In this paper, we provide a \textbf{FI}ne-grained \textbf{R}obust p\textbf{E}rformance \textbf{D}iagnosis (FIRED) framework. Our contributions can be summarized as follows: 
\begin{itemize}
\item We design an integrated framework to implement performance diagnosis effectively by putting metrics selection which can filter useless monitoring metrics, well-performed performance anomaly detection and prediction, and fine-grained root cause localization together.

\item We improve performance anomaly detection accuracy and robustness significantly by developing a novel deep ensemble method. The method can also predict anomalies in four minutes with F1 score higher than 0.8. 

\item We propose a real-time and fine-grained root cause localization pipeline by building dependency graph and executing random walk automatically. It can accurately localize the first root cause, and with the average localization accuracy higher than 0.7 of locating first four root causes.
\end{itemize} 

The rest of the paper is organized as follows. In section \ref{relw}, we review existing research about performance diagnosis and introduce anomaly detection and root cause localization methods. In section \ref{fram}, a performance diagnosis framework and its main components are introduced in detail. In section \ref{exp}, experiments and results for each component of our diagnosis framework are provided. Finally, we provide discussion and conclusion in section \ref{dis} and section \ref{con}. 

\section{Related works}
\label{relw}
Research about performance diagnosis is ongoing rapidly in clouds, e.g., micorservice\cite{wu2020microrca} and cloud datacenter\cite{roy2018cloud}. Ibidunmoye et al. \cite{ibidunmoye2015performance} reviewed performance anomaly detection and bottleneck identification methods, in which they formulated fundamental research problems, categorized detection methods, and proposed research trends and open challenges. In general, performance diagnosis frameworks include anomaly detection and root cause localization \cite{meng2020localizing}\cite{yu2021microrank}. This section will introduce related works in terms of performance anomaly detection and localization methods. 

For performance diagnosis, cloud application monitoring data needs to be preprocessed first. metrics selection is to select metrics related to run-time application status, and it can be used to reduce data dimension, improve detection accuracy and efficiency \cite{khalid2014survey}. In real scenarios, if fewer labels exist, feature selection methods, such as filter, wrapper, and embedded methods, can be used to choose a subset of all features \cite{chandrashekar2014survey}. In the situation that no labels exist, feature extraction methods that create a subset of new features from combination of existing features can be considered, such as PCA (Principal Components Analysis) \cite{yang2004two} and LDA (Linear Discriminant Analysis) \cite{wei2006lda}. This paper provides metrics selection methods and comparison experiments in section \ref{metr_ex} and \ref{mtc_eva}.

\subsection{Performance anomaly detection}
Machine learning-based anomaly detection methods can be reviewed based on supervised, semi-supervised, and unsupervised learning. Supervised learning methods have high accuracy \cite{hu2018detecting}, but it's not practical for application monitoring data because data labels are usually missing in reality and labeling data manually is time-consuming. Therefore, we mainly focus on semi-supervised and unsupervised machine learning methods. We will also specifically highlight ensemble learning methods. 

\begin{table}
\centering
\caption{\label{tab:pad} Unsupervised performance anomaly detection methods}
\resizebox{0.45\textwidth}{!}{
\begin{tabular}{lll}
\hline
Type & Algorithm & Description \\ \hline
\multirow{2}{*}{Density-based} & LOF\cite{breunig2000lof} & Local Outlier Factor\\ 
& LOCI\cite{papadimitriou2003loci} & Local Correlation Integral \\
\multirow{2}{*}{Distance-Based} & KNN\cite{ramaswamy2000efficient} & K Nearest Neighbors\\ 
& LDOF\cite{zhang2009new} & Local Distance-based Outlier Factor \\
Kernel-based & OCSVM\cite{scholkopf2001estimating} & One-Class Support Vector Machines\\
\multirow{2}{*}{Ensemble} & IForest\cite{liu2008isolation} & Isolation Forest\\
& Feature bagging\cite{lazarevic2005feature} & Subset of features are used \\
\multirow{2}{*}{Deep learning}  & AutoEncoder\cite{sakurada2014anomaly} & Fully connected AutoEncoder \\ 
& VAE\cite{kingma2013auto} & Variational AutoEncoder \\ 
\hline
\end{tabular}
}
\end{table}

\subsubsection{Anomaly detection methods}
Semi-supervised learning is developed under the situation where fewer labels exist. For example, Camacho \cite{camacho2019semi} et al. presented a semi-supervised approach for anomaly detection. The method extends the unsupervised multivariate statistical network monitoring approach based on principal component analysis (PCA) by introducing a supervised optimization technique to learn the optimum scaling in the input data. Experiments show that the semi-supervised method performs much better than unsupervised detection. 

Unsupervised learning methods are developed considering that there are usually no labels in reality. In table \ref{tab:pad}, we provide a classification of unsupervised performance anomaly detection methods. The table shows that density-based, distance-based, kernel-based, and tree-based methods are the most commonly used and usually focus on different features in the data. Therefore, the performance of these methods varies greatly for the data. Deep learning methods are also developing rapidly recently. For example, Su et al. \cite{su2019robust} provided a stochastic recurrent neural network named OmniAnomaly for multivariate time series anomaly detection in various devices. Deep learning methods are effective for large-scale datasets but are usually time-consuming to train.

Researchers usually focus on the data preprocessing phase to improve algorithm robustness for machine learning methods. For example, Bhagoji et al. \cite{bhagoji2018enhancing} propose the use of data transformations, including data dimension reduction via PCA, to enhance the resilience of machine learning methods. These researches show the importance of algorithm robustness, but more methods can be considered for improving algorithm robustness, such as ensemble learning. Research about anomaly prediction usually focuses on single-variable data and one-step prediction. For example, Wu et al. develop a prediction-driven anomaly detection scheme for single-variate time-series data. This paper provides the multi-step prediction based on multi-variate metrics for performance anomalies.

\subsubsection{Ensemble learning}
Ensemble learning is proposed to improve the accuracy and reduce the variance of an automated decision-making system \cite{galicia2019multi}. The primary assumption is that by combining several base models, the errors of a single model will likely be compensated by other models \cite{zhou2019ensemble}. For anomaly detection, the ensemble of anomaly scores by taking the maximum, and average actions \cite{aggarwal2015theoretical}. Research about ensemble learning can be reviewed based on supervised classification, semi-supervised and unsupervised clustering ensemble \cite{dong2020survey}.

Supervised ensemble learning needs to train with labels. Tama et al. \cite{tama2020enhanced} propose a stacked ensemble that uses three classifiers (random forest, gradient boosting machine, and XGBoost) and provides a generalized linear model (GLM) as a combiner. Semi-supervised ensemble learning mainly focuses on expanding the training set and utilizing the expanded training set to do classification or regression. Yu et al. \cite{yu2018multiobjective} proposed a multi-objective subspace selection process to generate the optimal combination of feature subspaces to improve the performance of the classifier ensemble. For unsupervised ensemble learning, research mainly focuses on consensus clustering, and includes pair-wise co-occurrence based methods\cite{fred2005combining}, graph partitioning based methods\cite{huang2015robust} and median partition-based methods \cite{huang2016ensemble}. 

In conclusion, weakly-supervised learning methods can be considered for performance anomaly detection because fewer labels exist in real scenarios. Existing detection methods rarely consider detection accuracy, algorithm robustness, and multi-step prediction simultaneously. In addition, ensemble learning methods are trying to improve detection accuracy by combining the characteristics of multiple models, but most of them are integrating detection methods linearly. Therefore, we can consider developing a detection method that integrates existing detection methods with non-linear ways, improving detection accuracy and robustness, and even predicting performance anomalies for cloud applications monitoring data.

\subsection{Root cause localization} 
In recent years, studies about localizing root causes for performance anomalies in cloud applications are developing. Researchers proposed machine learning, pattern recognition, and graph-based methods with data from cloud applications, such as logs, requests execution tracing data, and metrics to localize root causes \cite{notaro2021survey}. Graph-based methods can identify root causes and provide visibility of the issue with a visualized graph. Here, we review graph-based methods as follows. 

Graph-based methods identify the root causes by constructing a graph from observational data, then inferring and ranking the causes based on the graph. Service dependency graphs are usually built based on service deployment and service co-location. For example, Lin et al. \cite{lin2018microscope} provide the Microscope to builds the service causality graph based on network connection information. Wu et al. \cite{wu2020microrca} propose MicroRCA to build an attributed graph that models anomaly propagation across both services and machines. The service dependency graph can be used to localize root causes on services or containers, which are coarse-grained.

In addition, the casual graph for monitoring data can be constructed based on causality inference. 
Chen et al. provide \cite{chen2014causeinfer} Causeinfer to build the causal graph with the PC (named after the authors, Peter and Clark) algorithm based on the conditional independence test. Meng et al. propose \cite{meng2020localizing} MicroCause, which is a variant of the PC algorithm and considers the time-order of metrics to identify a causality graph of metrics. However, this method also has the time-lag limitations of monitoring data and has a higher computation burden than the PC algorithm. Therefore, most research for root cause localization is coarse-grained, and causal graphs to localize root causes in monitoring data are still developing. This paper will use the PC algorithm to build a causal graph for our monitoring data and evaluate the localization accuracy. 

In conclusion, we can implement the performance diagnosis framework to meet requirements in real scenarios. As for metrics selection, feature selection and feature extraction methods can be considered based on monitoring data. As for performance detection, we consider developing a novel detection method to improve detection accuracy and guarantee robustness based on machine learning methods and ensemble learning. Root cause localization based on causality inference and graph methods can be developed to discover causes in monitoring metrics with fine-grained.

\begin{figure*}[ht!]
\centering
\includegraphics[width=6.0in]{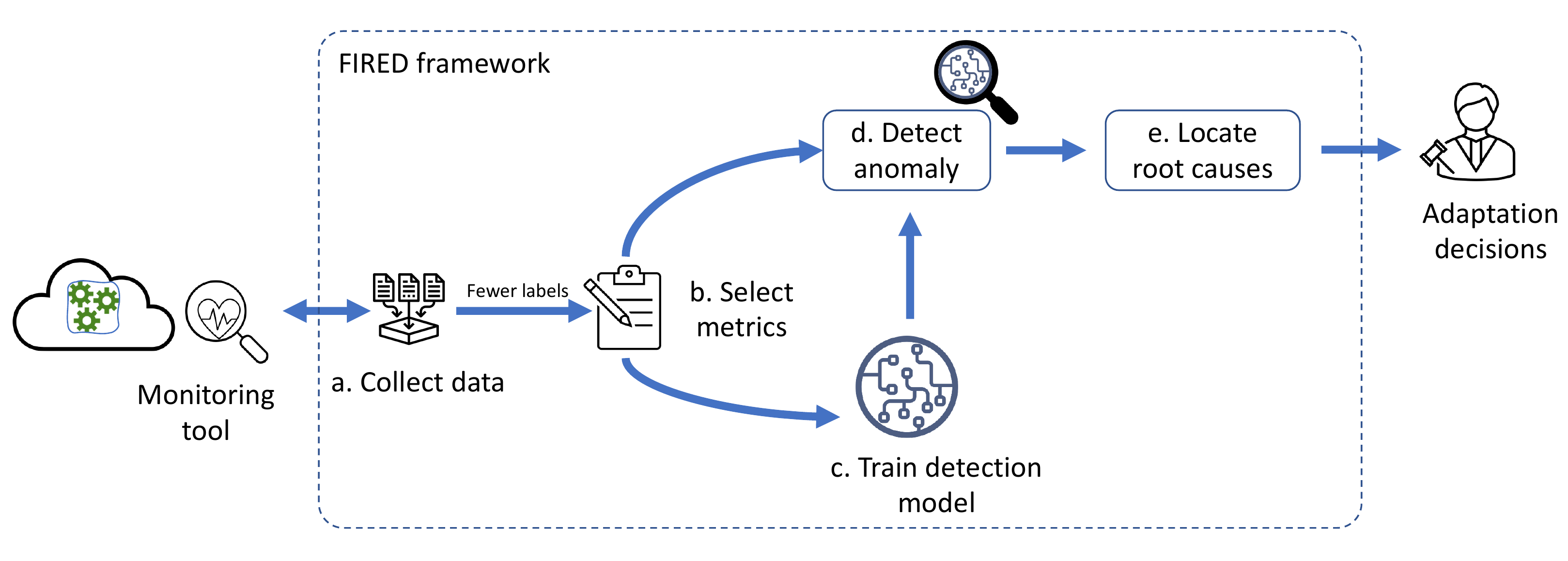}
\caption{The FIRED framework for cloud applications, which has model training and real-time testing. The framework includes \textbf{a}. collected data and fewer labels will be used for 
\textbf{b}. metrics selection and \textbf{c}. performance anomaly detection. Trained detection model will be used for \textbf{d}. real-time anomaly detection. Once an anomaly is detected, \textbf{e}. root cause localization will start to discover causes of the anomaly. The localization results can be used for recovery of cloud applications. }
\label{fig:workflow}
\end{figure*}

\section{Performance diagnosis framework}
\label{fram}
Based on related works, we can see that the three challenging requirement of cloud applications: good detection performance, multi-step prediction ability, real-time and fine-grained root cause localization, still need to be addressed. Therefore, a performance diagnosis framework which can effectively detect performance anomalies and localize root causes for cloud applications is needed. The effective performance anomaly detection should achieve high detection accuracy, good algorithm robustness, and multi-step prediction ability. The root cause localization should identify root causes with fine-grained, such as in metric granularity, with high accuracy. In this paper, we provide a \textbf{FI}ne-grained \textbf{R}obust p\textbf{E}rformance \textbf{D}iagnosis (FIRED) framework, which can be seen in Fig. \ref{fig:workflow}. The framework works with several steps. At first, (a) we collect multivariate time-series monitoring data continuously. In this paper, we mainly focus on system resource data, such as CPU and memory usage. In addition, fewer labels to indicate performance anomalies of an application will be used in our framework. For collected data, (b) the metrics selection will filter multivariate data with feature selection or extraction methods to select relevant metrics and reduce data dimensions. Subsequently, (c) selected metrics are used to train the performance anomaly detection method. We also use selected metrics of new data and then test them with the trained detection method. Once an anomaly occurs, (d) we start the root cause localization to discover the causes of the anomaly. The localization results can be used for rapid recovery of cloud applications. 

The diagnosis framework provides a deep ensemble method for performance anomaly detection that integrates existing methods based on ensemble learning. Considering that fewer labels exist in real scenarios, the deep ensemble method is a weakly-supervised learning method to improve detection accuracy, as described in \ref{method}. We evaluate the deep ensemble method for detection accuracy, robustness, and multi-step prediction ability compared with other detection methods. Furthermore, effective root cause localization enables rapid recovery for detected performance anomalies. We develop root cause localization based on causality inference and graph methods. We mainly evaluate localization accuracy, which will be introduced in \ref{rcl}. The main notations in this paper are given in table \ref{tab:notation}. 

\begin{table}[htb!]
\centering
\caption{\label{tab:notation} Notations and definitions} 
\begin{tabular}{|c|p{5.5cm}|}
\hline
\multicolumn{1}{|c|}{\textbf{Notion}} & \multicolumn{1}{c|}{\textbf{Definition}}
\\ \hline
$N$ & Number of all metrics \\ \hline
$n$ & Number of selected metrics \\ \hline 
$d$ & Number of samples in each metric \\ \hline
$k$ & Number of base learners \\ \hline
$K^t$ & Anomaly labels, $t$ is index of timestamps \\ \hline
$R_i^t$ & Collected monitoring data, $i$ is index of metrics, $t$ is index of timestamps. \\ \hline
$D_j^t$ & Data after metrics selection, $j$ is index of data dimensions, $t$ is index of timestamps. \\ \hline
$C_k^t$ & Anomaly scores vector of each base learner, $k$ is index of base learners, $t$ is index of timestamps.  \\ \hline
$O_k^t$  & Anomaly scores vector after normalization, $k$ is index of base learners, $t$ is index of timestamps. \\ \hline
$G$  & Dependency graph of metrics \\ \hline
\end{tabular}
\end{table}

\subsection{Metrics selection}
\label{metr_ex}
Multivariate data usually contains noise, introducing unnecessary variance into a developed model. Therefore, metrics selection to identify relevant metrics and reduce data dimension is needed. Considering fewer labels exist, we can easily select relevant metrics from multiple metrics with filter methods. Pearson’s correlation is generally used to measure the relevance between features, and it provides a fast estimation for feature selection. In addition, the feature extraction method PCA can be used to extract main features in data and reduce data dimensions without labels. We mainly focus on the feature selection method for application monitoring data in this paper. We also compare the performance of the feature selection and extraction methods. 


For original data, we apply z-score normalization \cite{saranya2013study} to ensure that all data have the same scale. The z-score method uses the mean and standard deviation of the original data for normalization so that the processed data follows the normal distribution. After normalization, we represent data with $R_i^t$ ($i=[1,...,N]$ is the index of resource metrics. $N$ is the number of all resource metrics. $t\in\mathbb{N^*}$ is the index of timestamps) as input data. Next, we provide the metrics selection for the input data. metrics selection has feature selection and extraction methods. After metrics selection, data $D_j^t$ will be used to diagnose the running status of cloud applications where $j=[1,...,n]$ is the index of data dimensions and $n$ is data dimensions after reduction.

With fewer labels, we can extract related system resource metrics automatically. We provide a filter method of correlation analysis for all metrics with these labels. For time-series data, we use $K^t$ to represent these labels, and $R_i^t$ represent resource metrics. We calculate the Pearson’s correlation of the labels with each resource metric. 

\begin{equation}
r_i = \frac{cov(K^t,R_i^t)}{\sigma_{K^t}\sigma_{R_i^t}}
\end{equation}

A significant test for the Pearson’s correlation.

\begin{equation}
t_i = r_i\sqrt{\frac{d-2}{1-r_i^2}}
\end{equation}

Here, $d$ is the number of timestamps, which is also the sample number of each resource metric. In order to filter out low correlation metrics, we set the threshold $t_i < 0.05$ and $|r_i| > 0.5$ for all correlation results. 

\begin{figure*}[ht!]
\centering
\includegraphics[width=6.2in]{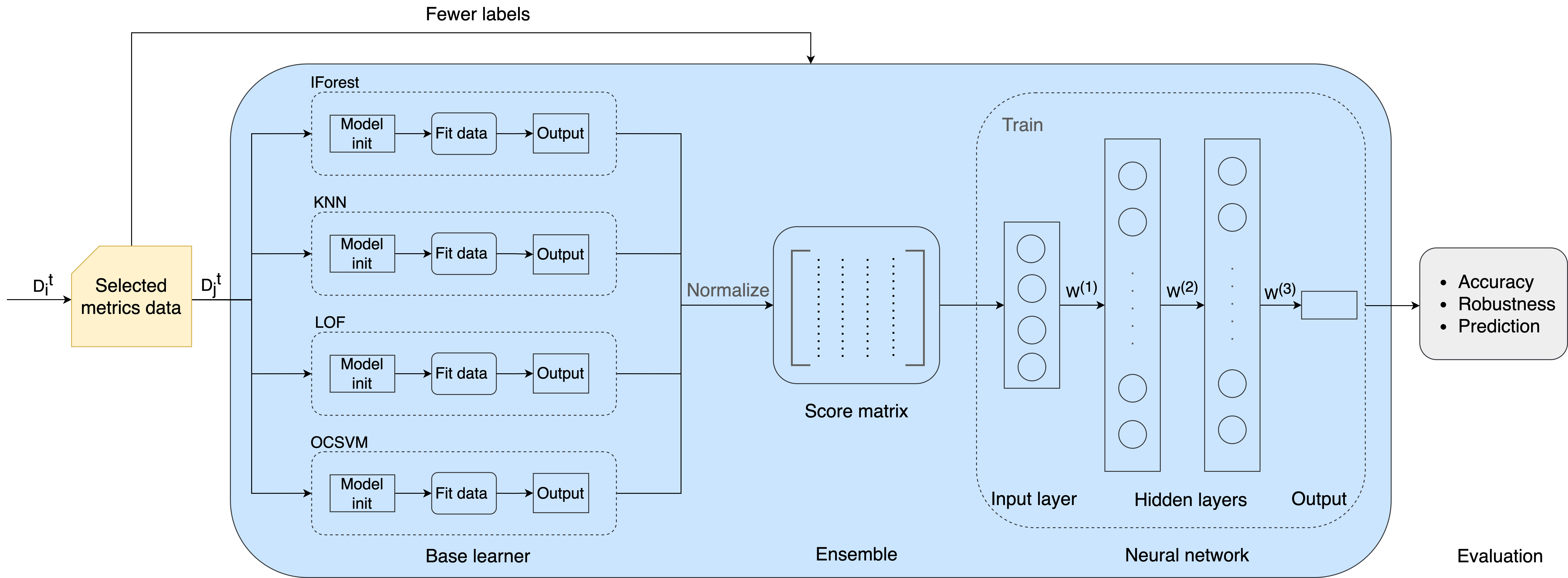}
\caption{The deep ensemble method for performance anomaly detection. }
\label{fig:algorithm}
\end{figure*}

In addition, we can use the feature extraction method PCA to transform a dataset with lots of variables into a smaller one that still contains most of the information in the original dataset. The process steps of PCA are: 1) getting the covariance matrix of original features; 2) calculating eigenvectors and eigenvalues of the covariance matrix to identify principal components; 3) sorting eigenvalues and selecting eigenvectors with high eigenvalues as feature vectors; 4) recasting the original data based on feature vectors. In step 3, the number of selected eigenvectors determines the data dimensions after reduction. Therefore, we can see that based on these calculations, PCA can be used without labels and achieves principal feature selection and data dimension reduction. In practice, we set the reduction dimension based on a calculated percentage of variance \cite{abdi2010principal}. We apply the correlation analysis and PCA to resource metrics $R_i^t$ and compare their performance in section \ref{mtc_eva}.

\subsection{Performance anomaly detection}
\label{method}

Lots of methods have been developed for performance anomaly detection, such as density-based, distance-based. However, these methods focus on different features in data and have diverse performances. Thus, it is reasonable to consider that integration of existing methods can extract more features from data and improve detection performance. Furthermore, ensemble learning has the assumption that by combining several base models, the errors of a single model will be compensated by others. However, the limitation of existing ensemble learning is that it tries to combine detection methods linearly. Therefore, in this paper, we provide a deep ensemble method that achieves the non-linear combination of existing detection methods, which can be seen in Fig.\ref{fig:algorithm}.

Because fewer labels exist in reality, we implement the deep ensemble method in a weakly-supervised manner.
We split preprocessed data $D_j^t$ into different amounts of labels as training data. The deep ensemble method includes three components, base learner, ensemble, and neural network. As base learners, we select four different unsupervised anomaly detection methods. Next, the outputs of these base learners are assembled into a score matrix. After that, the matrix is used as the input of a neural network for training. Each component will be introduced in detail next.   

\subsubsection{Base learner} 

Different anomaly detection methods usually focus on different features in data, such as density-based, distance-based, and this results in diverse performance on data. Therefore, to have a comprehensive understanding of monitoring data characteristics, we select four classic methods (IForest, KNN, LOF, OCSVM) from table \ref{tab:pad} as base learners. IForest is developed based on the decision tree algorithm \cite{freund1999alternating}. Many isolation trees make up an isolation forest to make anomaly detection results more credible. KNN is a distance-based algorithm \cite{ramaswamy2000efficient}, which calculates the distance (Euclidean, Manhattan) of points with neighbors as anomaly scores. LOF is a density-based algorithm \cite{breunig2000lof} and the density is compared with neighbors to determine anomalies. OCSVM is based on Support Vector Machine (SVM) \cite{scholkopf2001estimating}, which projects data through a kernel function into a high-dimensional space to classify them. 

For each base learner, the input is the preprocessed data $D_j^t$. The processing of input data includes model initialization, fitting data, and output anomaly scores, as shown in Fig. \ref{fig:algorithm}. Model initialization includes the setup of hyper-parameters, such as anomaly fractions, which can be set based on data characteristics. After fitting data, an anomaly score vector $C_k^t$ ($k$ represents the index of base detection methods) will be output. The anomaly score vector can be used to identify anomalies and evaluate the performance of each detection method. In our deep ensemble method, all anomaly score vectors from base learners are used as the input of ensemble methods, which will be introduced next. 

\subsubsection{Ensemble}

The outputs of base learners have different meanings and scales. For example, the anomaly score of IForest is calculated based on path depth, KNN is based on distance. Because all the features should be measured in the same units, we also apply z-score normalization \cite{saranya2013study} to ensure that all outputs have the same scale. After normalization, we can represent the anomaly scores vector $C_k^t$ of each base learner as $O_k^t$. Here, $k$ is the index of base detection methods and $k\in{[1,r]}$, $r$ is the number of base learners. Therefore, by taking each anomaly scores vector as the column, we can get the anomaly scores matrix $M$ as $$ M = 
 \left[
 \begin{matrix}
   O_1^1 & O_2^1 & O_3^1 & O_4^1 \\
   O_1^2 & O_2^2 & O_3^2 & O_4^2 \\
   \vdots & \vdots & \vdots & \vdots \\
   O_1^t & O_2^t & O_3^t & O_4^t \\
   \vdots & \vdots & \vdots & \vdots
  \end{matrix}
  \right] 
$$

For matrix $M$, classic ensemble learning methods exist, such as 1) maximum ensemble, 2) average ensemble and 3) weighted average ensemble. The maximum ensemble selects the max value of each row in matrix $M$ and forms a new anomaly score vector. The average ensemble calculates the average of each row in matrix $M$ and forms a new anomaly score vector. For the average ensemble, a limitation is that each base learner has an equal contribution to the final anomaly score. However, some learners perform better or worse than others. Therefore, weighted average ensemble which can assign different weights for base learners is developed. Based on the assumption that if a mixed model can maximize the information provided by each model, the mixed model has the best weights distribution strategy. Mutual information (MI) \cite{kuncheva2003measures} can measure the difference of models, which can be used to calculate the weight of each base learner. 

In table \ref{tab:enm}, we provide 5 samples as an example to show how maximum, average, and weighted average ensemble methods work. In the left part of the table, we show the anomaly scores of four base learners we chose for our comparison. In the right part, we can easily derive maximum and average anomaly scores. As for the weighted average ensemble, weights are assigned as (0.39, 0.28, 0.04, 0.29) for base learners based on the MI calculation. These new anomaly score vectors can be used to identify anomalies and evaluate the performance of these ensemble methods.

\begin{table}[htbp]
\centering
\caption{\label{tab:enm} Ensemble based methods example: on the left side is anomaly scores obtained by each base learner; on the right side is anomaly scores obtained through ensemble based methods}
\scalebox{0.8}{
  \begin{tabular}{ccccc|ccc}
  \hline
    Index & IForest & KNN & LOF & OCSVM & Max & Avg & Weighted Avg\\
    \hline
    1 & -0.41 & -0.23 & 0.14 & -0.88 & 0.14 & -0.35 & -0.49 \\
    2 & -0.18 & -0.03 & 0.63 & -0.86 & 0.63 & -0.11 & -0.33 \\
    3 & 2.29 & 5.14 & 1.07 & 0.62 & 5.14 & 2.28 & 2.76 \\
    4 & 2.36 & 4.56 & 0.86 & 0.11 & 4.56 & 1.97 & 2.42 \\
    5 & 1.99 & 1.5 & -0.3 & -0.19 & 1.99 & 0.75 & 1.14 \\
  \hline
\end{tabular}
}
\end{table}

\subsubsection{Deep neural network}

The ensemble learning methods above try to combine different anomaly scores linearly. However, linear combination is inadequate to represent the information extracted by each model well. Therefore, we combine base learners through a nonlinear way with a neural network in the deep ensemble method. MLP (multi layer perceptron) is a supplement of a feed-forward neural network, which consists of the input layer, output layer, and hidden layer. An MLP is suitable for classification or regression problems where inputs are assigned a class or real-value label. Therefore, the deep ensemble method is weakly-supervised, and it needs to train with fewer labels. In addition, the MLP can be replaced by other deep learning methods, like LSTM \cite{graves2005framewise}, CNN \cite{krizhevsky2012imagenet}. We also evaluate the feasibility of the replacement in our experiments. 

The MLP architecture can be seen in Fig. \ref{fig:algorithm}. The input layer receives the anomaly score matrix $M$ at first. We have 2 hidden layers consisting of an arbitrary number of neurons and use ReLU as an activation function. The output layer has one neuron and outputs the probability using the softmax activation function. We define $x=[O_1^t, O_2^t, O_3^t, O_4^t]$. $W^{(1)}$ and $b^{(1)}$ are weights and biases of the first layer. $W^{(2)}$, $b^{(2)}$ and $W^{(3)}$, $b^{(3)}$ are weights and bias of the two hidden layers. The output can be calculated based on the below functions. 

\begin{equation}
\begin{split}
    z^{(1)}&=W^{(1)}x + b^{(1)}, \\ 
    h^{(1)}&=ReLu(z^{(1)}), \\
    z^{(2)}&=W^{(2)}h^{(1)} + b^{(2)}, \\
    h^{(2)}&=ReLu(z^{(2)}), \\
    z^{(3)}&=W^{(3)}h^{(2)} + b^{(3)}, \\
    h^{(3)}&=softmax(z^{(3)})
\end{split} 
\end{equation}

For the output $h^{(3)}$, we can calculate the difference between the predicted result and actual result $y$ with the cross-entropy error function below. Here, $y$ is the label at time $t$. The optimization goal is to minimize this equation by constantly adjusting parameters.
\begin{equation}
    l = -y^T\log{h^{(3)}}
\end{equation}

The deep ensemble method needs to train with fewer labels, and then the trained model can be applied to other data to detect anomalies. If we let $y$ be the label of time $t+s$ ($s$ is steps), we can train a model with multi-step prediction ability. Experiments to evaluate the performance of the deep ensemble method and the impact of different amount of labels can be seen in section \ref{ad_res}.

\subsection{Root cause localization}
\label{rcl}
\begin{figure}[ht!]
\centering
\includegraphics[width=3.4in]{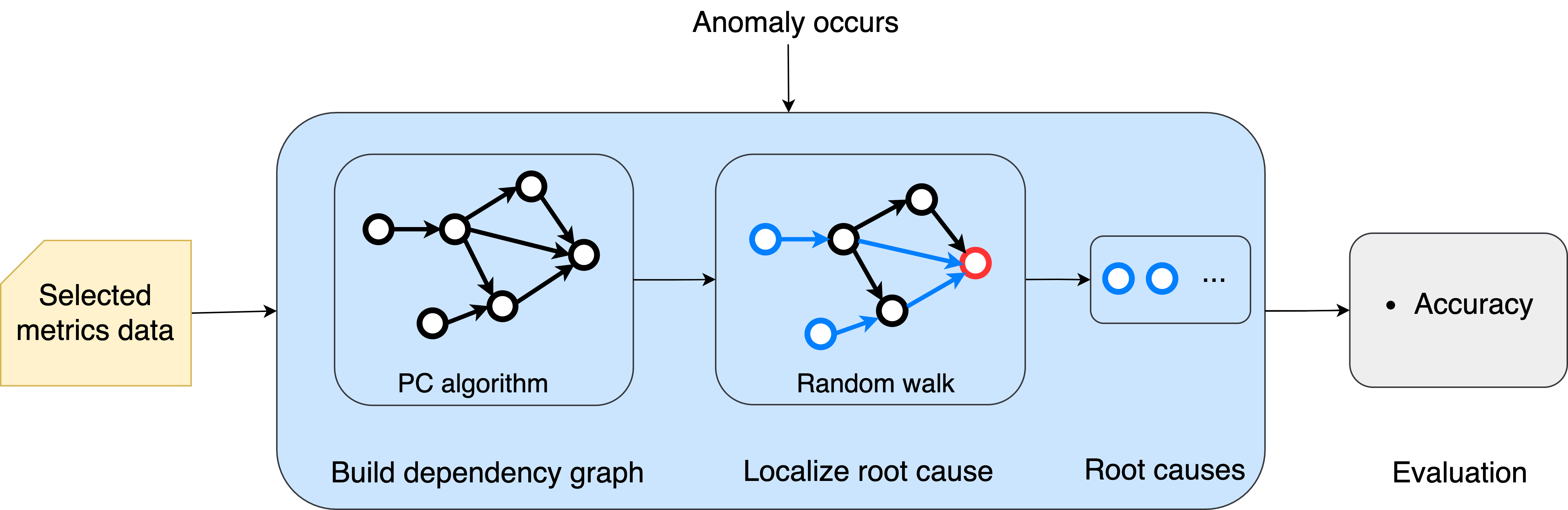}
\caption{\label{fig:rcl-wf} The root cause localization pipeline}
\end{figure}

Performance anomaly detection allows us to know the status of cloud applications. When an anomaly occurs, localizing the root causes of the anomaly can enable the application to recover effectively. We provide the pipeline of root cause localization for performance anomalies in Fig. \ref{fig:rcl-wf}. The input data consists of selected metrics and anomaly labels. Selected metrics are data after feature selection, which can be identified as CPU related, memory related. While data after feature extraction can't be used because there is no clear meaning of extracted features. For these selected time-series metrics, we extract their causal relations and build a dependency graph with the PC algorithm. Based on the dependency graph, we use a random walk to find the propagation path and localize root causes. Finally, we evaluate the localization accuracy. 

\subsubsection{Build dependency graph}
The causality between system resources and application performance is obvious, for example, low network bandwidth will cause high response latency. To extract the relation, causal graph is commonly used in practical applications because of its intuitiveness. The most popular method for constructing a causal graph from observational data is the PC algorithm \cite{kalisch2007estimating}. 

We use the PC algorithm to discover the causal relationship between system resources and performance anomalies. There are four steps to build a dependency graph with the PC algorithm:
\begin{itemize}
    \item Construct a fully connected graph of the $m$ random variables (all nodes are connected).
    \item Perform a conditional independence test on each adjacent variable under the significance level $\alpha$. If conditional independence exists, the edge between the two variables is removed. In this step, the size of the conditional variable set S increases step by step until no more variables can be added into S.
    \item Determine the direction of some edges based on v-structure\cite{neuberg2003causality}.
    \item Determine the direction of the rest of the edges.
\end{itemize}

Based on the PC algorithm, we build a dependency graph for all selected metrics $D_j^t$ and the anomaly labels $K^t$. We define the anomaly labels $K^t$ as an anomaly indicator. In addition, other causality inference methods can be used to build the dependency graph, like ANM \cite{hoyer2008nonlinear}. We also compare their localization performance in our experiments. 

\subsubsection{Localize root causes}
In a dependency graph, there can be many paths that point to the anomaly indicator, which makes it hard to localize root causes. To solve this problem, we apply a Random Walk algorithm to the dependency graph, which performs well in capturing anomaly propagation. The random walk procedure in a dependency graph is presented in Algorithm 1.

\begin{algorithm}
\caption{Random walk for DAG}\label{algorithm}
\SetKwInOut{Input}{Input}\SetKwInOut{Output}{Output}
	
\Input{DAG $G$, path length $l$, start node $N$} 
\Output{Path points to the start node}
\BlankLine 

path = [N]

\While{len(path)\textless{l}}{cur\_node = path[-1]\;
\eIf{len(list(G.predecessors(cur\_node)))\textgreater{0}}{predecessor = random.sample(list(G.predecessors(cur\_node, 1)\;
path.extend(predecessor)\;}
{break\;}}
\Return path
\end{algorithm}

In this algorithm, we set the anomaly indicator as the start node. Furthermore, we end up with a path pointing to the start node by randomly selecting the predecessors of the current node. We iterate the algorithm many times and get several paths. The last node of each path can be regarded as the root cause. By counting and ranking root cause nodes, we can finally get the root cause set. 

\begin{figure}[ht!]
\centering
\includegraphics[width=3.4in]{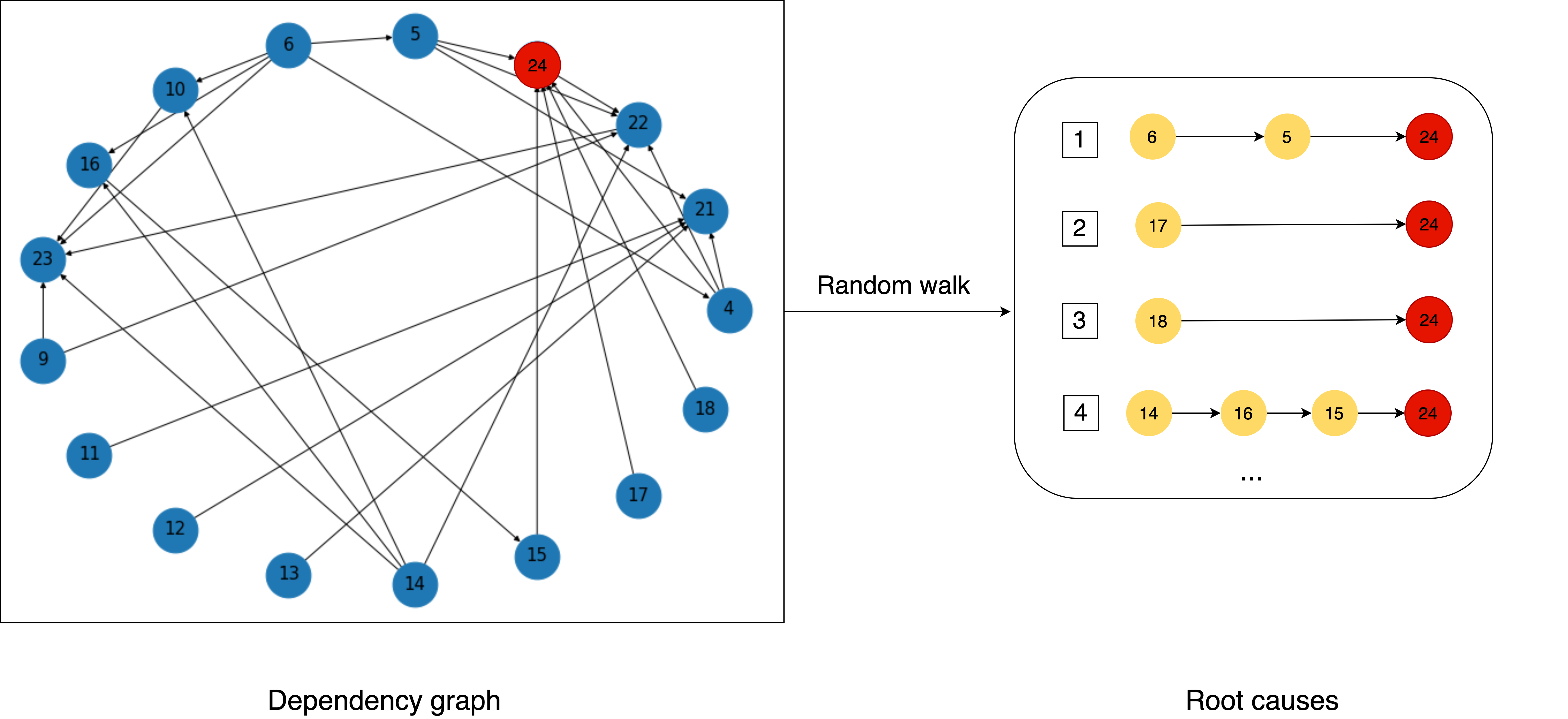}
\caption{\label{fig:cg_plot} Visualization of root cause localization pipeline}
\end{figure}

We provide an example for the root cause localization pipeline in Fig.\ref{fig:cg_plot}. Nodes 0-23 represent selected metrics, and node 24 represents the anomaly indicator. We first build a dependency graph for these metrics with the PC algorithm. Isolated nodes which have no causality relation with others are removed in the dependency graph. Next, for the dependency graph, we use the random walk algorithm to get paths pointing to the anomaly indicator and rank all root cause nodes. We can see that there are paths like $6 \rightarrow 5 \rightarrow 24$, $17 \rightarrow 24$, $18 \rightarrow 24$, $14 \rightarrow 16 \rightarrow 15 \rightarrow 24$. After ranking, the root causes are localized as \{6, 17, 18, 14\}. 

\section{Evaluation and results}
\label{exp}
We provide different experiments to evaluate each component in the FIRED framework. 
\begin{itemize}
    \item We evaluate metrics selection by comparing the detection performance of base learners with data processed by metrics selection methods. 
    \item We conduct the deep ensemble method experiments in terms of detection accuracy, algorithm robustness, multi-step prediction ability and the effect of different amounts of data.
    \item We check the feasibility of the root cause localization pipeline, compare different causality graph methods, and observe time spent. 
\end{itemize}
In this section, we will introduce two datasets that are used in our experiments first. For the evaluation of each component, we present the experimental settings and evaluation results in detail next. 

\subsection{Datasets}
We use a decentralized application (DApp) monitoring data and a public dataset in our experiments.

\subsubsection{DApp monitoring data}

\begin{figure}[ht!]
\centering
\includegraphics[width=2.6in]{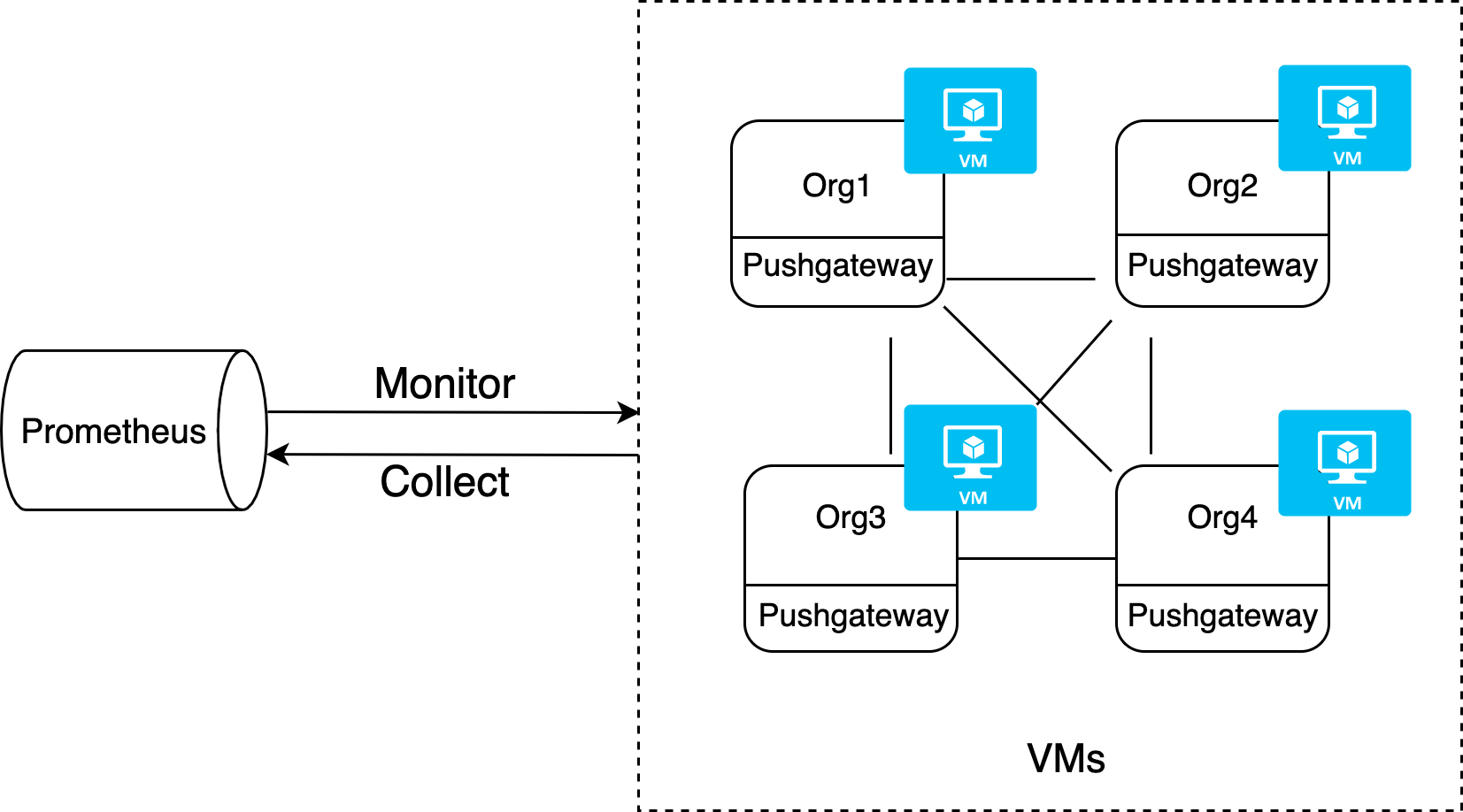}
\caption{Monitoring a Hyperledger Fabric based Decentralized application and continually collecting data with Prometheus}
\label{fig:monitor}
\end{figure}

In business scenarios where real-time transactions are required, e.g., energy trading or crowd journalisms \footnote{Example use cases are from the EU ARTICONF project, www.articonf.eu}, the quality of service (QoS) metrics of a DApp such as transaction throughput, latency, and failure rates are critical to the business value. To deliver such a quality critical DApp in cloud environments, one needs to select cloud services carefully, customize their capacities, and monitor the run-time status of the application. Fig. \ref{fig:monitor} shows a DApp example developed with Hyperledger Fabric\footnote{https://www.hyperledger.org/use/fabric}. For the DApp, different organizations, which contain many peer nodes, are deployed on different cloud infrastructure services (VMs), and they are monitored by a tool Prometheus \footnote{https://prometheus.io/}. We also use Prometheus to collect real-time data and use Caliper\footnote{https://www.hyperledger.org/use/caliper} to simulate workload generation. 

For a running DApp, different types of monitoring data can be collected, such as application-level data like workloads, blockchain-level data like transaction numbers, and system-level data like resource usages. System-level data can reflect application performance anomalies directly. For example, CPU overload can cause high response delay. For a DApp, resource usage like high CPU/MEM usage may cause transaction failure. Therefore, we mainly focus on system resource metrics, which can be seen in table \ref{tab:metrics}. When the DApp receives transaction requests stably, we add system pressures with \textit{stress-ng}\footnote{https://kernel.ubuntu.com/~cking/tarballs/stress-ng/}, such as I/O pressure to inject anomalies manually. We increase I/O pressure for 20 minutes every hour. We keep monitoring the DApp for 12 hours and collect data every 15 seconds, resulting in 3237 samples and 229 resource-related metrics for our experiments. In addition, an important metric that represents the number of transaction failures can be seen as the anomaly indicator of the DApp. We use this metric to select relevant resource metrics. 

\begin{table}[htb!]
\caption{\label{tab:metrics} Description of resource metrics} 
\begin{tabular}{|c|p{5.5cm}|}
\hline
\multicolumn{1}{|c|}{\textbf{Resource Metrics}} & \multicolumn{1}{c|}{\textbf{Description}}\\ \hline
CPU related       & Per core and overall load, usage, idle time, I/O wait time, hard and soft interrupt counts, context switch count, etc. \\ \hline
Memory related    & Free, cached, active, inactive, dirty memory, etc. \\ \hline
Disk related      & Disk space used, IOps, I/O usage, read/write rate, etc. \\ \hline
Network related   & Receive/transmit network traffic, etc.    \\ \hline
\end{tabular}

\end{table}

\subsubsection{Public SMD data}

The SMD (Server Machine Dataset) is a dataset collected and made publicly available by a large internet company\cite{su2019robust}. It contains data from many different server machines, and each one can collect 38 monitoring metrics. Each metric has an index rather than a specific name. Domain experts have labeled anomalies in the SMD based on incident reports. To evaluate the performance of anomaly detection methods in this paper, the information of SMD data we use in our experiments can be seen in table \ref{tab:dataInfo}.

\begin{table}[htb!]
\centering
\caption{\label{tab:dataInfo} General information of the two datasets}
\resizebox{0.45\textwidth}{!}{
\begin{tabular}{|c|c|c|c|c|}
\hline
\textbf{Dataset} &  
\textbf{\makecell[c]{Number of \\ samples}} &
\textbf{\makecell[c]{Number of \\ features}} & 
\textbf{\makecell[c]{Number of \\ selected features}} & 
\textbf{\makecell[c]{Anomaly \\ fraction (\%)}} \\ \hline
DApp monitoring data & 3237 & 229 & 24 & 28.14  \\ \hline
SMD data & 28479 & 38 & 5 & 9.46  \\ \hline
\end{tabular}}
\end{table}

For evaluating the three components in the performance diagnosis framework, we only use the DApp monitoring data for the metrics selection evaluation and root cause localization because we know the detailed information of all metrics. As for performance anomaly detection, we use both the DApp monitoring data and the SMD data to evaluate the performance of the deep ensemble method. 

\subsection{metrics selection evaluation}
\label{mtc_eva}
We execute experiments to validate the metrics selection component in the performance diagnosis framework. We only use the DApp monitoring data here because we know the detailed information of each metric, for example, CPU usage, system load. The metrics selected by the two methods -- correlation analysis and PCA, and all metrics are used as the input to the four base learners. Afterward, we check the effect of metrics selection methods by comparing the detection performances of the four base learners.

\subsubsection{Experimental setting}

The DApp monitoring data is collected from a deployed DApp in a cloud environment. Here, we use Azure\footnote{https://azure.microsoft.com/en-us/} as the cloud environment and deploy the monitor component and DApp separately. The monitor component is deployed on a VM with the following properties: Ubuntu 18.04 as operating system, 2CPU, 4G Memory, 32GiB Storage. The DApp is deployed on VMs which have properties: Ubuntu 18.04 as the operating system, 4CPU, 16G Memory, and 32GiB Storage.

For the feature selection method, correlation analysis, we set the threshold $t_i < 0.05$ and $|r_i| > 0.5$ to filter low correlation metrics. For feature extraction, we need to determine the reduction dimensions of PCA. In general, PCA needs to retain as much variance information of original data as possible, such as 95\%. Therefore, we set the reduction dimensions to $15$ for DApp monitoring data based on a calculated percentage of variance \cite{abdi2010principal}. 

We evaluate the detection performance of these base learners with F1 score to indicate accuracy, and time spent to indicate efficiency. F1 score is a function of both Precision and Recall. The Precision is about how much of the data detected as anomalies are true anomalies, while recall is about how much of the real anomaly data is detected as anomalies. So, we calcuate F1 score as below:  
\begin{equation}
F1\ score = 2*\frac{Precision*Recall}{Precision+Recall}
\end{equation}

\subsubsection{Evaluation results}

\begin{figure}[ht!]
\centering
\includegraphics[width=3.5in]{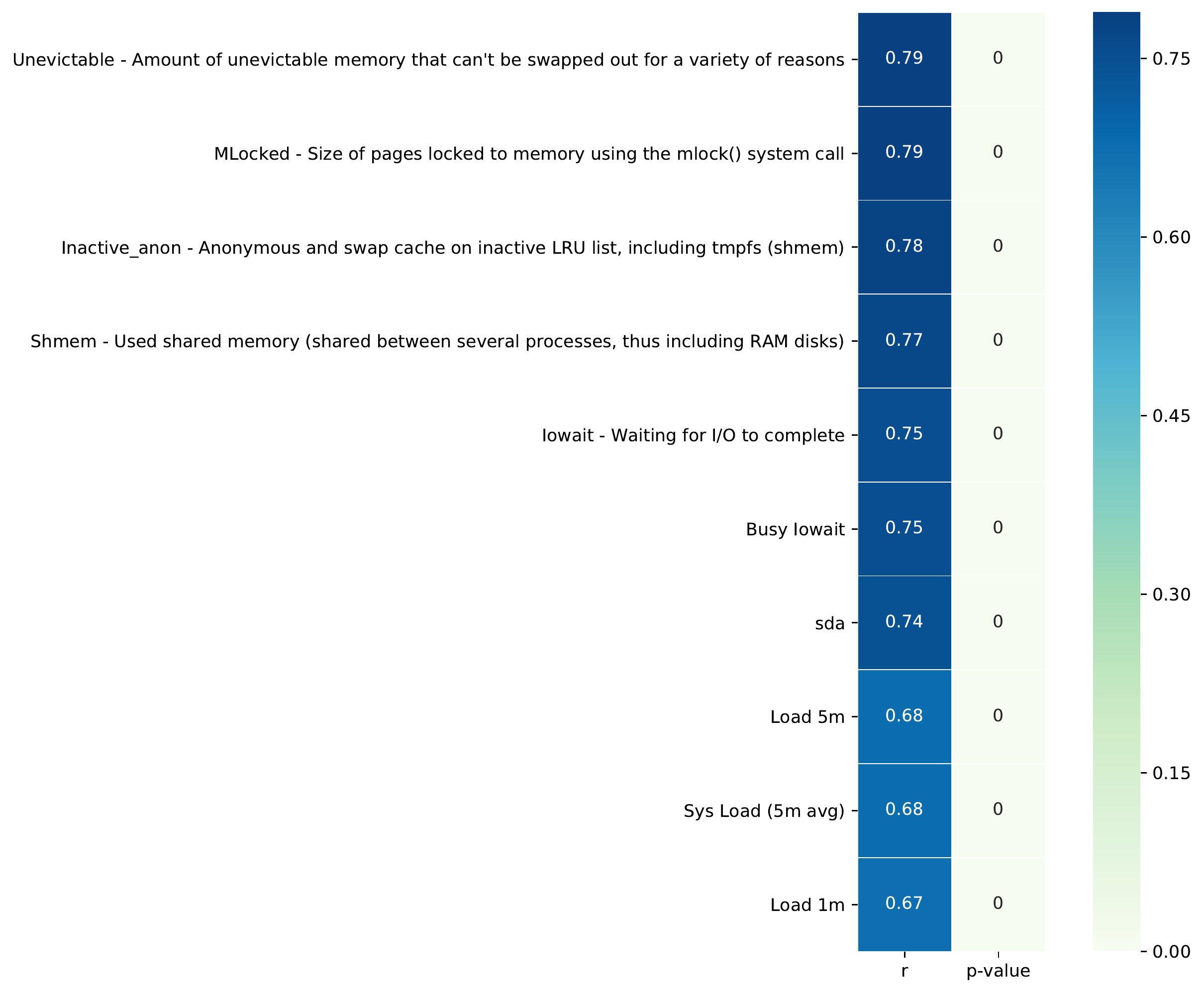}
\caption{Top 10 resource metrics with high relevance to performance anomalies}
\label{fig:corr}
\end{figure}

For metrics selection, we apply the correlation analysis and PCA separately on the DApp monitoring data. Fig.\ref{fig:corr} shows the correlation analysis result based on r-values in descending order. We calculate the correlation between all monitoring metrics and fewer labels. The results show that metrics like the amount of unevictable memory, iowait have high r-values, which means there is a relationship between these metrics and the occurrence of the transaction failure anomaly. In addition, we reduce data dimensions from 229 to 15 based on PCA method. 

\begin{figure}[ht!]
\centering
\includegraphics[width=3.4in]{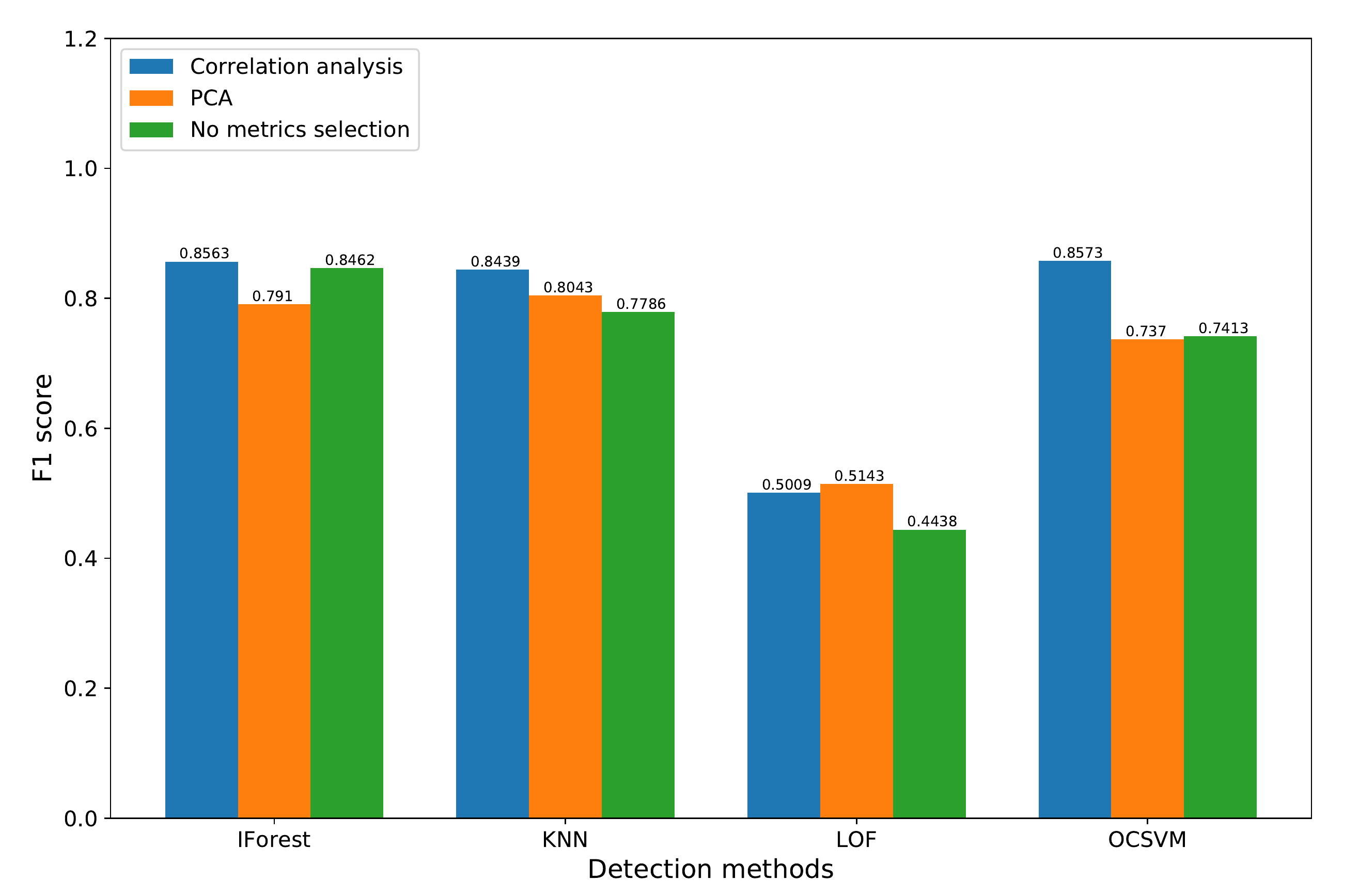}
\caption{Detection accuracy by base learners as function of the metrics selection methods: correlation analysis, PCA and no metrics selection.}
\label{fig:perf_metric}
\end{figure}

\begin{figure}[ht!]
\centering
\includegraphics[width=3.4in]{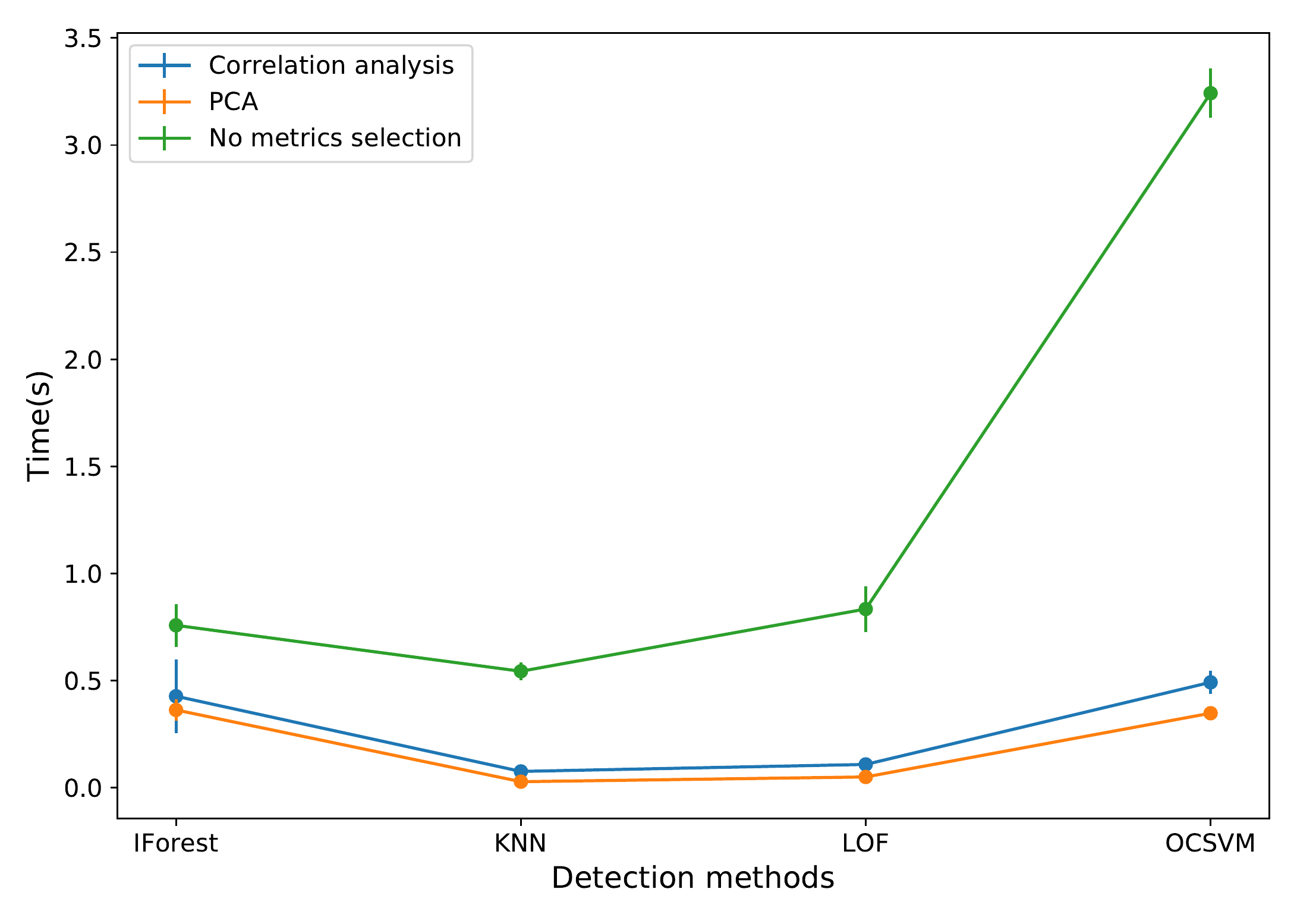}
\caption{Time spent by base learners as function of the metrics selection methods: correlation analysis, PCA and no metrics selection.}
\label{fig:time_metric}
\end{figure}

We then compare the effects of metrics selection methods based on the performance of base learners. We use the data after correlation analysis, PCA, and without metrics selection as the input of base learners, respectively. The F1 score and time spent of each base learner can be seen in Fig.\ref{fig:perf_metric} and Fig.\ref{fig:time_metric}. In Fig.\ref{fig:perf_metric}, we can see that after correlation analysis, three base learners, IForest, KNN and OCSVM have the highest F1 score. In comparison, the F1 score of LOF is slightly lower. For PCA, the F1 scores of KNN and LOF are higher than without metrics selection. But the F1 scores of IForest and OCSVM are lower than without metrics selection. We can say that metrics selection based on correlation analysis improve the detection accuracy. In addition, in Fig.\ref{fig:time_metric}, we can see that without metrics selection, the time spent is about 2 to 10 times for each base learner compared with using metrics selection.  

In conclusion, we can see that with metrics selection, the detection accuracy is improved and the time spent is reduced compared with no metrics selection. In addition, correlation analysis has better detection accuracy than PCA. Next, we will use data after metrics selection as the input of performance anomaly detection and root cause localization. 
\subsection{Performance anomaly detection evaluation}
\label{pad_eva}

We execute experiments to validate the performance anomaly detection in terms of detection accuracy, algorithm robustness, and multi-step prediction ability. Our experiments are implemented based on the DApp monitoring data and the SMD data.  


\subsubsection{Experimental setting}
We design several experiments to evaluate detection performance of the deep ensemble method. We use the DApp monitoring data after correlation analysis based on the metrics selection experiments above as the input of the deep ensemble method. Because SMD data after correlation analysis has only 1 metric left, which remove too much information, we use the SMD data after PCA as the input. The first experiment is to compare the detection accuracy of the deep ensemble method with base learners and linear ensemble methods. Because the deep ensemble method uses fewer labels to train the model, it is unfair to compare its results with these unsupervised methods. We then combine each base learner with an MLP to train the model and compare its performance with the deep ensemble method. In addition, for the deep ensemble method, we also try to find out the impact of the amounts of labels for training a model, considering there are only fewer labels for monitoring data in reality. 

The performance of these detection models is evaluated from three aspects: accuracy, robustness, and multi-step prediction ability. We still focus on the F1 score for detection accuracy. Our experiments present the time spent of each unsupervised detection method and test time of the deep ensemble method. For robustness, we test detection methods on the two different datasets, and we rank detection results to represent performance consistency, which can clearly show the detection performance comparison. To calculate the robustness, we take the average rank of detection methods on the two datasets. Moreover, we normalize the rank with the calculation: 
\begin{equation}
Robustness\ score = \frac{Rank-Rank_{max}}{Rank_{min}-Rank_{max}}
\end{equation}
Here, $Rank_{max}$ is the maximum of rank numbers, and $Rank_{min}$ is the minimum of rank numbers. We evaluate prediction ability with accuracy, which is also represented by the F1 score. 

As for the base learners, their hyper-parameters are set as below. Anomaly fractions of the two datasets need to be determined at first. For the DApp monitoring data, we inject anomalies for 20 minutes every hour, so the anomaly fraction is about 0.3. For SMD data, we use the default anomaly fraction, which is 0.1. Next, the hyper-parameters of each base learner need to be determined. For IForest, we define the tree number as 100. In KNN, the neighbor number is 5. In LOF, we set the neighbor number as 20. And in OCSVM, we use the RBF (radial basis function) kernel function.  

There is no hyper-parameter for max, average, and weighted average ensemble methods. For the deep ensemble method, we do the train/test split at first. We use 50\% of labels to train the model. We also compare the performance of training with different amount of labels (10\%, 30\%, 50\%, 70\%, 90\%). Next, we need to determine the hyper-parameters in the MLP, which are the same for the two datasets. The input layer has 4 neurons because we have 4 base learners. In addition, we set 20 neurons in the two hidden layers and the output layer as 1. We train 100 epochs and set the batch size as 20. We use the Adam optimizer for stochastic gradient descent with an initial learning rate of $10^{-3}$ during model training. We also trained the deep ensemble method 10 times. We show the error bar in our figures and take the average of evaluation metrics in our tables, such as F1 score and times, as the final result. 

\subsubsection{Performance anomaly detection results} 
\label{ad_res}

\begin{table}
\centering
\caption{\label{tab:comp_all_d} Comparison of different detection methods on the DApp monitoring data}
\resizebox{0.5\textwidth}{!}{
\begin{tabular}{ |c|c|c|c|c|c|}
\hline
Method & Precision & Recall & F1 Score & Rank 
&  Time(s) \\ 
\hline
IForest & 0.8418 & 0.8802 & 0.8563 & 5 &
0.4274$\pm$0.1715 \\ 
\hline
KNN & 0.8358 & 0.854 & 0.8439 & 6 &
0.0764$\pm$0.0106 \\ 
\hline
LOF & 0.5175 & 0.5215 & 0.5009 & 8 &
0.109$\pm$0.0161 \\ 
\hline
OCSVM & 0.8869 & 0.8374 & 0.8573 & 4 &
0.4923$\pm$0.0532 \\ 
\hline
Ensemble\_max & 0.8238 & 0.8516 & 0.8352 & 7 & 
1.0543$\pm$0.1348 \\
\hline
Ensemble\_avg & 0.8539 & 0.861 & 0.8602 & 2 &
1.0542$\pm$0.1348 \\
\hline
Ensemble\_w\_avg & 0.8589 & 0.8608 & 0.8598 & 3 & 
1.1007$\pm$0.1342 \\
\hline
Deep ensemble & \textbf{0.8996} & \textbf{0.8861} & \textbf{0.8923}$\pm$0.0076 & \textbf{1} &
1.6869$\pm$0.0219 \\
\hline
\end{tabular}}
\end{table}

\begin{table}
\centering
\caption{\label{tab:comp_all_s} Comparison of different anomaly detection methods on the SMD data}
\resizebox{0.45\textwidth}{!}{
\begin{tabular}{ |c|c|c|c|c|c|}
\hline
Method & Precision & Recall & F1 Score & Rank &
Time(s) \\
\hline
IForest & 0.7131 & 0.8266 & 0.7515 & 2 &
1.2763$\pm$0.0074 \\ 
\hline
KNN & 0.5905 & 0.741 & 0.5713 & 7 &
0.2896$\pm$0.0063 \\ 
\hline
LOF & 0.5425 & 0.5642 & 0.5468 & 8 &
0.5092$\pm$0.0098 \\
\hline
OCSVM & 0.6126 & 0.7945 & 0.6047 & 6 &
25.1878$\pm$0.644 \\
\hline
Ensemble\_max & 0.6705 & 0.8417 & 0.7058 & 3 & 
26.1372$\pm$0.9133 \\
\hline
Ensemble\_avg & 0.6451 & 0.8279 & 0.6676 & 5 & 
26.1368$\pm$0.9133 \\
\hline
Ensemble\_w\_avg & 0.6599 & 0.8377 & 0.6907 & 4 & 
26.2498$\pm$0.9182 \\ 
\hline
Deep ensemble & \textbf{0.8293} & \textbf{0.7821} & \textbf{0.8025}$\pm$0.0101 & \textbf{1} &
28.8262$\pm$0.0117\\
\hline
\end{tabular}}
\end{table}

\textbf{A. comparison with unsupervised detection methods.} In this experiment, we compare detection accuracy and algorithm robustness of the deep ensemble method with base learners and linear ensemble methods on both the DApp monitoring data and SMD data. 

The table \ref{tab:comp_all_d} shows the result for DApp monitoring data. We can see that IForest, KNN, and OCSVM have quite high F1 score and OCSVM with a value of 0.8573, while LOF is the lowest, which demonstrates that these base learners have performance inconsistency. Ensemble methods improve their detection accuracy by combining features extracted by these base learners. For linear ensemble methods, we can see that the F1 score of the average ensemble is 0.8602, which is slightly higher than base learners. The deep ensemble is developed based on ensemble methods, and it has the highest F1 score, 0.8923. The result shows that deep ensemble provides a significant improvement for detection accuracy. As for the time spent on these methods, we can see that time spent for base learners is less than 0.5s, of which KNN and LOF only use about 0.1s, while ensemble methods take about 1s. For the deep ensemble, the time spent is about 1.69s, which is higher than other methods mainly because the calculation is based on base learners. 

Table \ref{tab:comp_all_s} shows the result for SMD data. We can see that the F1 score of the IForest is 0.7515, which is higher than other base learners. The performance of linear ensemble methods is not good enough because base learners affect these ensemble methods while other base learners (KNN, LOF, and OCSVM) perform poorly. The most important observation is that the deep ensemble has the best F1 score, 0.8025, which is much higher than other methods. As for the time spent by these methods, we can see that the OCSVM has high time spent because the kernel function calculation of large-scale data is time consuming. Linear ensemble methods and the deep ensemble method are based on the processing of all base learners. So, the maximum, average, and weighted average ensemble spend about 26.2s, which is high than base learners. The deep ensemble spends about 28.8s, which is higher than other methods also because of OCSVM and the computational cost of the neural network. 

As for algorithm robustness, we provide the rank results in table \ref{tab:rank}. We rank the detection accuracy of all detection methods. We can see that the deep ensemble method has the best detection accuracy for both the DApp monitoring data and SMD data. Linear ensemble methods have good robustness compared with base learners. In contrast, base detection methods show performance inconsistency, for example, the IForest performs good for SMD data but bad for the DApp monitoring data. We can see that the deep ensemble method improves both detection accuracy and algorithm robustness. 

\begin{table}
\centering
\caption{\label{tab:rank} Rank result of algorithm robustness}
\resizebox{0.45\textwidth}{!}{
\begin{tabular}{|c|c|c|c|c|c|c|c|c|}
\hline
Method & IForest & KNN & LOF & OCSVM & \makecell[c]{Emsemble\\ \_max} & 
\makecell[c]{Ensemble\\ \_avg} & 
\makecell[c]{Ensemble\\ \_w\_avg} & 
\makecell[c]{\textbf{Deep}\\ \textbf{\_ensemble}} \\ 
\hline
\makecell[c]{DApp monitoring \\ data} & 5 & 6 & 8 & 4 & 7 & 2 & 3 & \textbf{1} \\
\hline
SMD & 2 & 7 & 8 & 6 & 3 & 5 & 4 & \textbf{1}\\ 
\hline
Avg rank & 3.5 & 6.5 & 8 & 5 & 5 & 3.5 & 3.5 & \textbf{1} \\
\hline
Robustness score & 0.6429 & 0.2143 & 0 & 0.4286 & 0.4286 & 0.6429 & 0.6429 & \textbf{1} \\
\hline
\end{tabular}}
\end{table}

\begin{figure}[h]
\centering 
\begin{minipage}[b]{0.45\textwidth} 
\centering 
\includegraphics[width=0.9\textwidth]{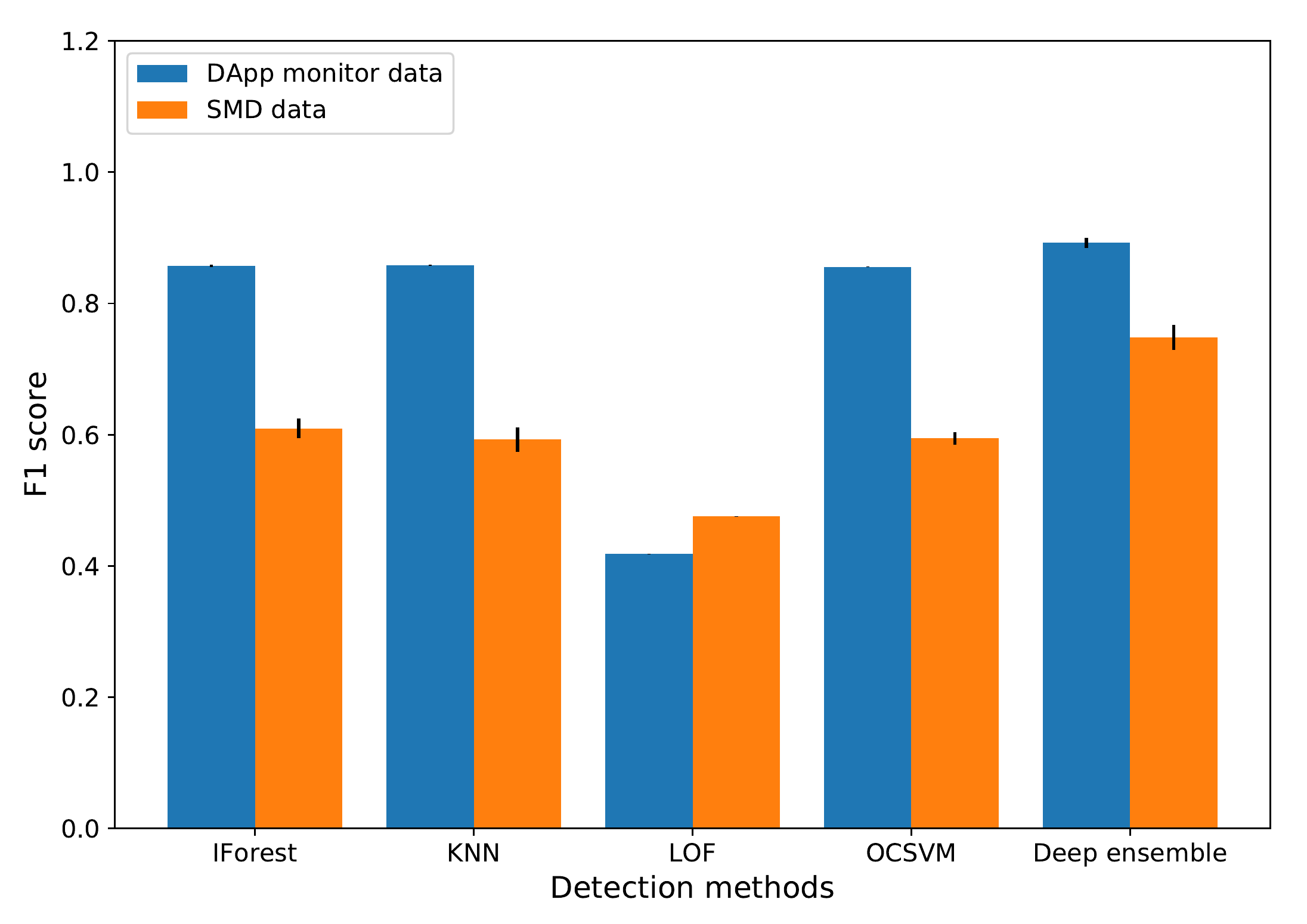}
\caption{\label{fig:base_mlp} Detection accuracy of deep detection methods}
\end{minipage}
\begin{minipage}[b]{0.5\textwidth} 
\centering 
\includegraphics[width=0.9\textwidth]{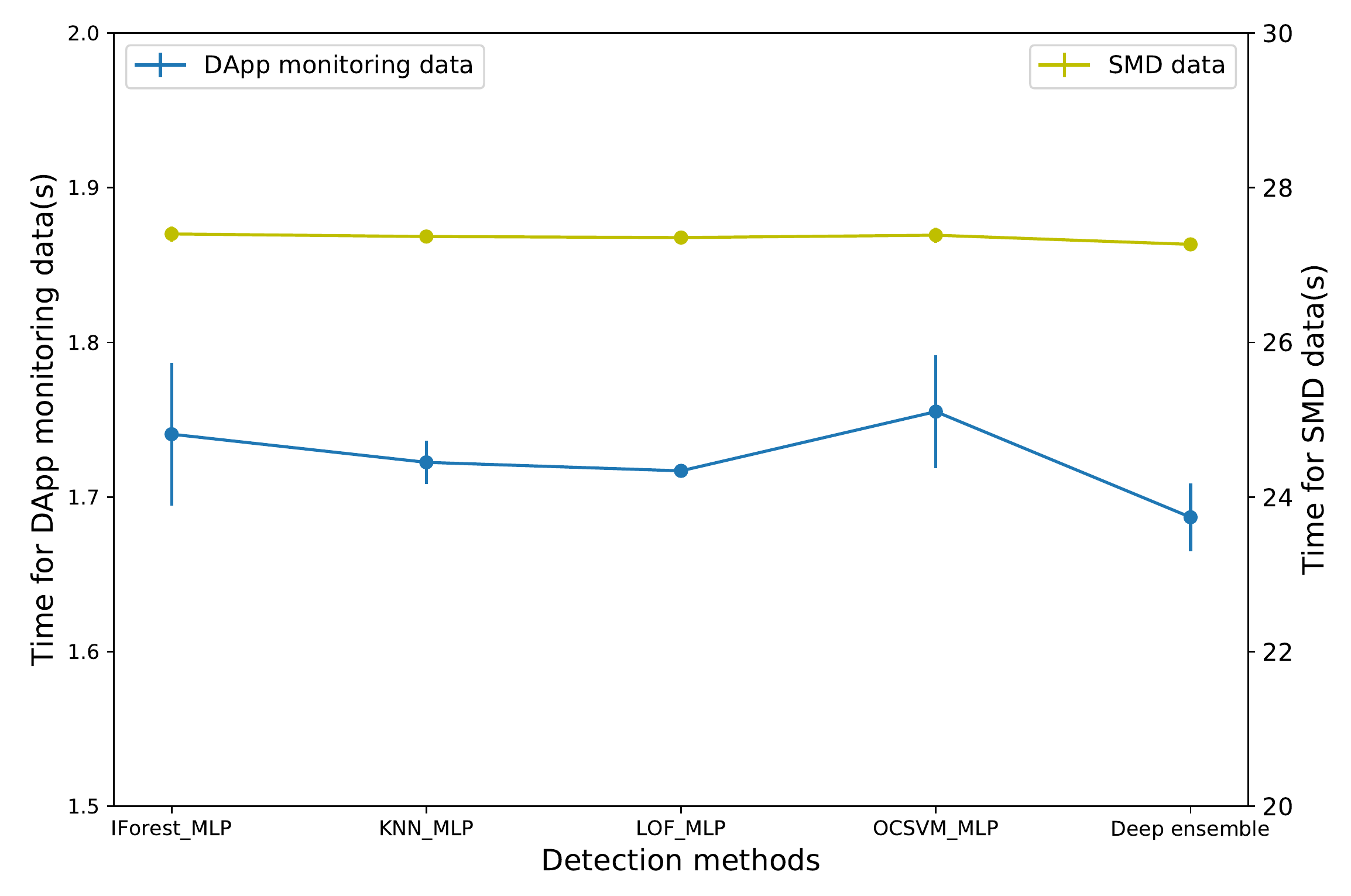}
\caption{\label{fig:time_base_mlp} Time spent of deep detection  methods}
\end{minipage}
\end{figure}

\begin{table}[htb!]
\centering
\caption{\label{tab:labels_d} Performance of different deep ensemble methods}
\resizebox{0.45\textwidth}{!}{
\begin{tabular}{ |c|c|c|c|c| } 
\hline
\multirow{2}*{Detection methods} & 
\multicolumn{2}{c|}{DApp monitoring data} &
\multicolumn{2}{c|}{SMD data} \\
\cline{2-5}
~ & F1 score & Time(s) & F1 score & Time(s) \\ 
\hline
Deep ensemble (MLP) & 0.8923$\pm$0.0076 & 1.0887$\pm$0.0219 &  0.8025$\pm$0.0101 & 26.2683$\pm$0.0217 \\
\hline
Deep ensemble (CNN) & 0.8839$\pm$0.0063 & 1.1434$\pm$0.0913 &  0.7765$\pm$0.0127 &  32.4284$\pm$0.1025 \\
\hline
Deep ensemble (LSTM) & 0.8936$\pm$0.0031 & 1.3831$\pm$0.1192 &  0.8196$\pm$0.0087 & 32.8886$\pm$0.3167 \\
\hline
\end{tabular}}
\end{table}

\textbf{B. comparison with other deep detection methods.} 
The deep ensemble method needs fewer labels to train, making it unfair to compare it with unsupervised methods. Therefore, we design experiments to compare its performance with other weakly-supervised detection methods. We create deep detection models by combining each base learner with an MLP and comparing their performance with the deep ensemble method at first. In addition, we also extend the deep ensemble method by replacing MLP with CNN and LSTM, and we provide a comparison of their detection performance.

We show the comparison of detection accuracy in figure \ref{fig:base_mlp}. We can see that the deep ensemble method has the highest F1 score for both the DApp monitoring data and SMD data, 0.8923 and 0.8025, respectively. The result shows that the deep ensemble method achieves the best detection accuracy, and it is a stable method that can be applied to different datasets. In figure \ref{fig:time_base_mlp}, we show the test time for each method. We can see that for the DApp monitoring data, the test time of all methods is similar and small, around 1.7s. For the SMD data, the test time for all methods is similar but quite high, about 27s, still because of time-consuming OCSVM kernal function calculation and computational cost of the neural network. 

As for replacing MLP with CNN and LSTM, the comparison of their detection performance can be seen in table \ref{tab:labels_d}. We can see that for both the DApp monitoring data and SMD data, the deep ensemble (LSTM) has the highest F1 score. This is reasonable because LSTM can extract long-term dependencies in data which is suitable for time-series data. The deep ensemble (CNN) has the lowest F1 score because the pooling layer in CNN will compress information, and this does not suit time-series data very well. As for spent time, we can see that these deep learning methods take similar time for test data. In contrast, the deep ensemble (MLP) method takes less time because of less computation. While the deep ensemble (CNN) is slightly faster than LSTM this is because we use fewer parameters in CNN. Therefore, we can see that the deep ensemble method can be extended easily, and replacing the MLP with LSTM can improve detection accuracy. 

In conclusion, we can see that the deep ensemble method has superior detection performance compared with other deep detection methods. In addition, the deep ensemble method the deep ensemble (LSTM) has the best detection accuracy. 

\begin{figure}[ht!]
\centering
\includegraphics[width=3.4in]{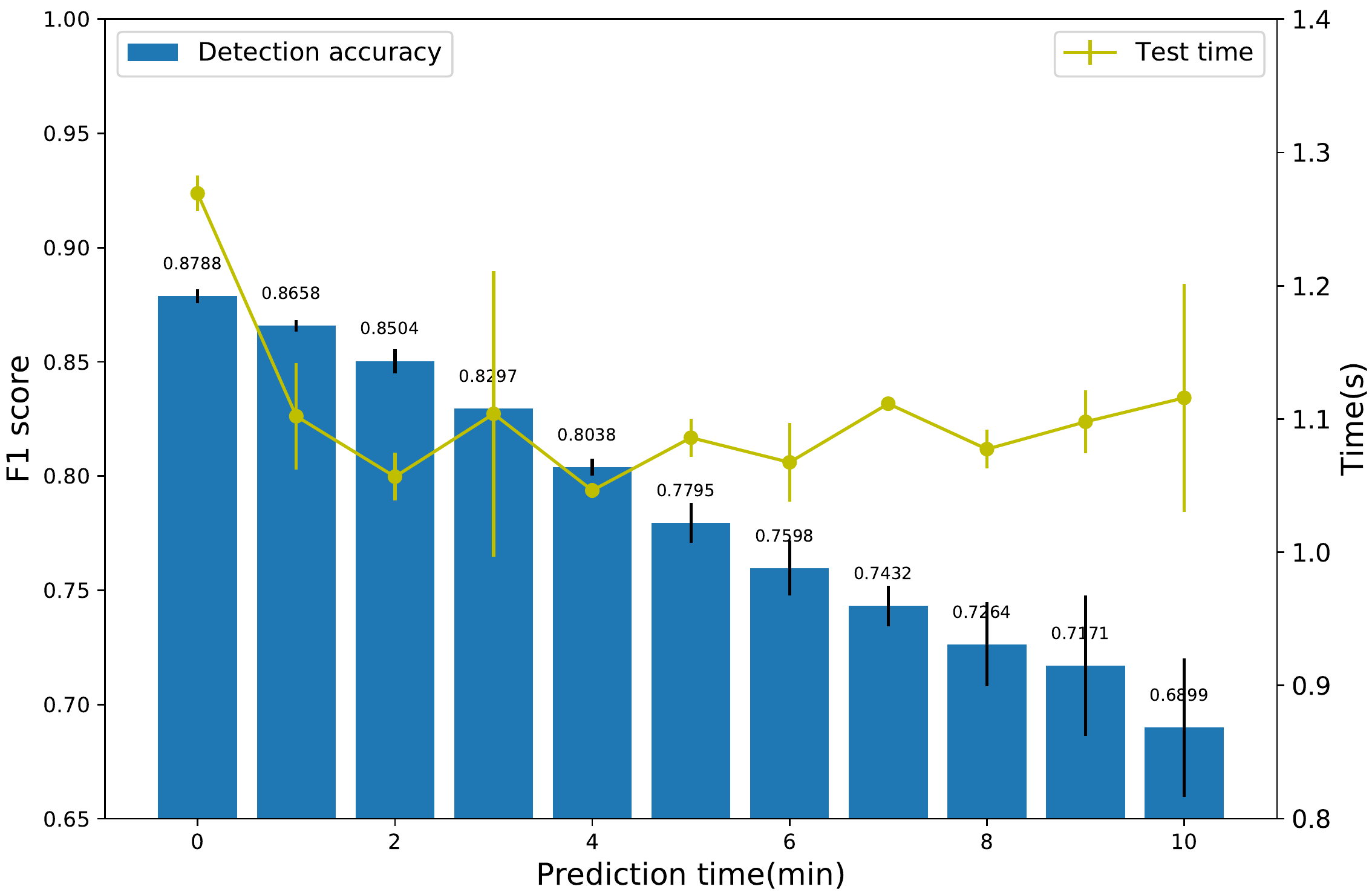}
\caption{Prediction accuracy and time spent for different time steps on the DApp monitoring data}
\label{fig:pred_time_d}
\end{figure}

\textbf{C. multi-step prediction ability.} 
With the deep ensemble method, we can predict multi-step performance anomalies. We test the prediction ability of the deep ensemble method on the DApp monitoring data. We collect the DApp monitoring data with 15s time interval, so we use every 4 steps, which is 1min as the prediction step. We predict that the anomaly will happen or not after one, two, or three minutes. To evaluate the prediction ability, we present the prediction accuracy in Fig. \ref{fig:pred_time_d}.

In Fig. \ref{fig:pred_time_d}, we can see that the longer the prediction time, the lower the detection accuracy, which means that it is difficult to predict long-term anomalies. Also, we can see that within four minutes, all F1 scores are higher than 0.8, which is pretty good. Therefore, we can predict anomalies in the next four minutes. We also show the time spent for testing the prediction ability in Fig. \ref{fig:pred_time_d}. We can see that the testing time is around 1.1s, which is not high. In conclusion, with the deep ensemble method, we can predict performance anomalies, and the prediction can have high accuracy in four minutes. 


\begin{table}[htb!]
\centering
\caption{\label{tab:labels_d} Impact of amounts of labels on the DApp monitoring data and SMD data}
\resizebox{0.45\textwidth}{!}{
\begin{tabular}{ |c|c|c|c|c| } 
\hline
\multirow{2}*{Number of labels} & 
\multicolumn{2}{c|}{DApp monitoring data} &
\multicolumn{2}{c|}{SMD data} \\
\cline{2-5}
~ & F1 score & Time(s) & F1 score & Time(s) \\ 
\hline
10\% labels & 0.8796$\pm$0.0042 & 1.6796$\pm$0.016 & 0.7729$\pm$0.0092 & 25.3878$\pm$0.0103 \\
\hline
30\% labels & 0.8881$\pm$0.0046 & 1.6797$\pm$0.0087 & 0.7902$\pm$0.0163 & 25.4847$\pm$0.1181 \\
\hline
50\% labels & 0.8923$\pm$0.0076 & 1.6869$\pm$0.0219 & 0.8025$\pm$0.0101 & 26.2683$\pm$0.0217 \\
\hline
70\% labels & 0.8948$\pm$0.005 & 1.6868$\pm$0.0227 & 0.8115$\pm$0.0088 & 26.2883$\pm$0.0262 \\
\hline
90\% labels & 0.8968$\pm$0.0047 & 1.6957$\pm$0.0326 &  0.8142$\pm$0.0149 & 26.3515$\pm$0.1824 \\
\hline
\end{tabular}}
\end{table}

\textbf{D. impact of different amounts of labels.} The deep ensemble needs to train with fewer labels. We use 50\% labels for all the experiments above. Here, we design an experiment to test the impact of  amounts of labels to evaluate the detection ability of the deep ensemble method. We also conduct experiments on both the DApp monitoring data and SMD data, and the results can be seen in table \ref{tab:labels_d}.

For the DApp monitoring data and SMD data, we set different amounts of labels (10\%, 30\%, 50\%, 70\%, 90\%) to train the deep ensemble method. We can see that with only 10\% labels for training and testing for all data, the F1 score of the deep ensemble method is higher than all the base learners and linear ensemble methods. In addition, more labels are used for training, and the F1 score is higher, which is easy to explain given that more samples with labels provide more information to learn. Besides, the time spent for each dataset in the table is similar, and this shows that the amounts of labels for training the model have little effect on the test time. To conclude, the deep ensemble method can achieve superior performance with fewer labels, such as 10\%, to train, and the trained models can be used on other data with high detection accuracy. 

In conclusion, our experiments validate the performance anomaly detection in the FIRED diagnosis framework. We show that the deep ensemble method achieves the best detection accuracy and algorithm robustness compared with other detection methods. Also, the deep ensemble method can predict anomalies in four minutes with F1 score higher than 0.8. In addition, as a weakly-supervised learning method, the deep ensemble method can get superior detection performance with fewer labels, such as 10\% amounts of labels.

\subsection{Root cause localization evaluation}
\label{rcl_eva}
We conduct experiments to validate the feasibility of root cause localization in the performance diagnosis framework. Our experiments are implemented based on DApps monitoring data, because we have clear description of each metric. We will identify which metrics in the DApps monitoring data cause performance anomalies. 

\subsubsection{Experimental setting}
We apply root cause localization methods on the DApps monitoring data. In table \ref{tab:metrics}, we provide 24 selected metrics and 1 anomaly indicator of DApp monitoring data. We classify them into CPU/MEM/NET/Disk related metrics. For the DApp we are monitoring, we add I/O pressure to inject anomalies. The root causes are I/O related as shown in table \ref{tab:metrics}. As for methods in the localization pipeline, we set $\alpha$ in the PC algorithm as 0.05, and the iteration for the random walk is 500. 

\begin{table}[htb!]
\centering
\caption{\label{tab:metrics} Description of selected resource metrics} 
\begin{tabular}{|c|c|p{3.2cm}|c|}
\hline
\multicolumn{1}{|c|}{\textbf{Index}} & \multicolumn{1}{c|}{\textbf{Type}} &
\multicolumn{1}{c|}{\textbf{Metric}} &
\makecell[c]{Ground\\ truth}
\\ \hline
0 & Memory related   &   Unevictable - Amount of unevictable memory that can't be swapped out for a variety of reasons & \\ \hline
1 & Memory related   &  Size of pages locked to memory using the mlock() system call & \\ \hline
2 & Memory related   & Inactive\_anon - Anonymous and swap cache on inactive LRU list, including tmpfs (shmem) & \\ \hline
3 & Memory related   & Shmem - Used shared memory (shared between several processes, thus including RAM disks) &  \\ \hline
4 & CPU related  &  Iowait - Waiting for I/O to complete & \checkmark \\ \hline
5 & CPU related  & Busy Iowait  & \checkmark \\ \hline
6 & Disk related & sda  & \checkmark \\ \hline
7 & CPU related  & Load 5m   &\\ \hline
8 & CPU related  & Sys Load (5m avg)  &\\ \hline
9 & CPU related  &  Load 1m  &\\ \hline
10 & CPU related &  CPU Busy  &\\ \hline
11 & Memory related & Pagesout - Page out operations   & \\ \hline
12 & Disk related &  sda - Successfully written bytes  & \checkmark \\ \hline
13 & Disk related &  sda - Written bytes  & \checkmark  \\ \hline
14 & Disk related &  sda - discard  & \checkmark \\ \hline
15 & Network related & OutOctets - Sent octets  &  \\ \hline
16 & Network related & trans eth0   & \\ \hline
17 & Disk related &  Processes blocked waiting for I/O to complete  & \checkmark \\ \hline
18 & Memory related &  Dirty - Memory which is waiting to get written back to the disk  &   \\ \hline
19 & CPU related & Sys Load (15m avg)   & \\ \hline
20 & CPU related & Load 15m   & \\ \hline
21 & Memory related & Writeback - Memory which is actively being written back to disk   &  \\ \hline
22 & Disk related & sda - Writes completed   & \checkmark \\ \hline
23 & CPU related &  Idle  & \\ \hline
24 & Anomaly indicator &  txn\_fail\_label  & \\ \hline
\end{tabular}
\end{table}

To evaluate the accuracy of root cause localization, we use two performance metrics: $AC@k$ and $Avg$. These two metrics are most commonly used to evaluate the rank result of the root cause localization task \cite{meng2020localizing}\cite{wu2020microrca}. $AC@k$ represents the probability that top $k$ results localized by algorithms include the real root causes for a given anomaly. When the $k$ is small, the higher $AC@k$ indicates the algorithm identifies the actual root cause more accurately. We calculate $AC@k$ as follows:

\begin{equation}
    AC@k = \sum_{i<k}\frac{R[i]\in{V_{rc}}}{min(k,|V_{rc}|)}
\end{equation}

where $R[i]$ is the result of rank of all metrics for the anomaly. $V_{rc}$ is the root cause set of the anomaly. $Avg$ evaluates the overall performance of the localization algorithm by computing the average $AC@k$. The calculation is as follows:
\begin{equation}
    Avg = \frac{1}{k}\sum_{1\leq{j}\leq{k}}{AC@j}
\end{equation}

\subsubsection{Root cause localization results}

\begin{figure}[ht!]
\centering
\includegraphics[width=3.6in]{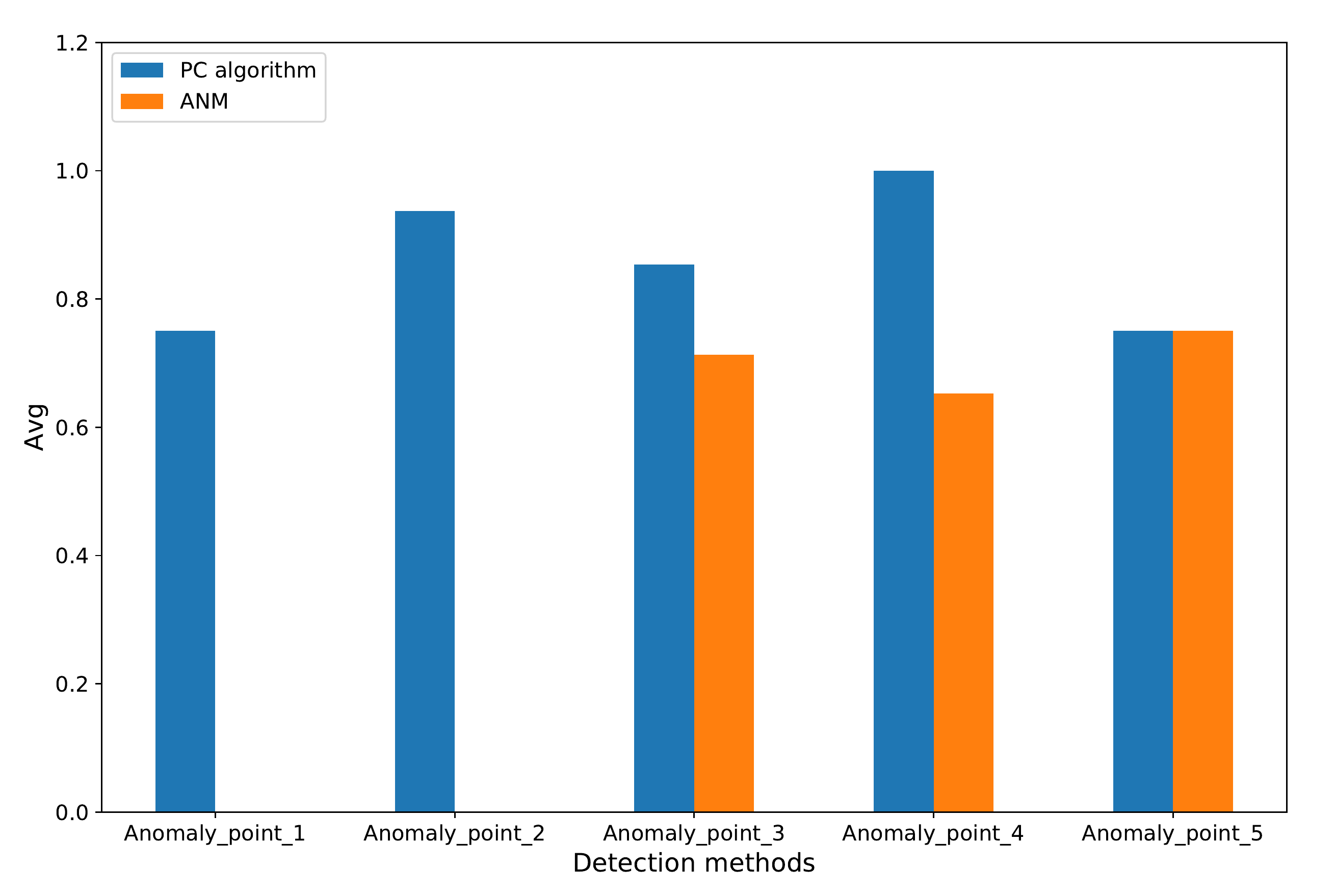}
\caption{\label{fig:compare} root cause localization accuracy based on different dependency graph building methods}
\end{figure}

When an anomaly is detected, we start to localize its root causes. With anomaly injection, we get several anomalies periods, and each of them lasts for 20 minutes. We select 5 of them randomly and compare the localization accuracy based on two different dependency graph building methods: the PC and the ANM algorithms. The comparison results can be seen in Fig. \ref{fig:compare}. We use $Avg$ to evaluate localization accuracy. We can see that, for different anomaly points, the PC algorithm has better performance than the ANM algorithm. In addition, for anomaly\_point\_1 and anomaly\_point\_2, we can see that the ANM algorithm does not discover real root causes because the dependency graph does not extract causality relations from data. Therefore, the PC algorithm has better localization accuracy and more stable performance for the DApp monitoring data. We also show detailed localization performance of PC algorithm in table \ref{tab:loc_acc}. We calculate the detection accuracy of AC@1, AC@2, AC@3, AC@4, and Avg, and present the localization results. We can see that the real root cause can be localized directly for different anomaly points. As for anomaly\_point\_2 and anomaly\_point\_3, we can see that multiple root causes are discovered, including many real root causes, so the localization accuracy is high. Also, we provide the time spent of building dependency graph and localizing root causes, we can see that the localization for these anomalies can complete within 1s, which means that the localization can be done in real-time with given data. 

In conclusion, our experiments demonstrate the feasibility of real-time root cause localization in the performance diagnosis framework. In addition, we build the dependency graph and localize root causes at fine-grained. Our experiments also show that the localization based on the PC algorithm is accurate for the DApp monitoring data. 

\begin{table}
\centering
\caption{\label{tab:loc_acc} root cause localization accuracy for anomalies with the PC algorithm}
\resizebox{0.5\textwidth}{!}{
\begin{tabular}{ |c|c|c|c|c|c|c|}
\hline
Metric & AC@1 & AC@2 & AC@3 & AC@4 & Avg & Time(s)\\
\hline
Anomaly\_point\_1 & 1 & 0.5 & - & - & 0.75 & 0.843\\ 
\hline
Anomaly\_point\_2 & 1 & 1 & 1 & 0.75  & 0.9375 & 0.797 \\ 
\hline
Anomaly\_point\_3 & 1 & 0.5 & 0.667 & 0.75 & 0.729 & 0.591\\ 
\hline
Anomaly\_point\_4 & 1 & - & - & - & 1 & 0.688 \\ 
\hline
Anomaly\_point\_5 & 1 & 0.5 & - & - & 0.75 & 0.913\\ 
\hline
\end{tabular}}
\end{table}

\section{Discussion}
\label{dis}
This paper introduces the FIRED framework for performance diagnosis, including metrics selection, performance anomaly detection, and root cause localization. We design a number of experiments to validate the FIRED framework. However, for the methods and experiments in this paper, some aspects can still be improved. 

As for metrics selection, we will consider combining correlation analysis and PCA to extract useful information in data. We provide a deep ensemble method for performance anomaly detection and evaluate it in the section \ref{pad_eva}. The performance of the deep ensemble method is severely affected by base learners because the outputs of base learners are assembled as the inputs of subsequent processing. In this paper, we select the four base detection methods manually based on their differences. Next, we can consider automatically selecting suitable base learners based on data distributions. In addition, for the deep ensemble (LSTM) method, which has the best detection accuracy, we will conduct hyperparameter tuning to optimize its performance and explore approaches to improve detection efficiency. In the root cause localization pipeline, we will develop advanced methods to build the dependency graph, like based on Graph Autoencoder\cite{ng2019graph}, which can extract non-linear relations between metrics to improve localization accuracy in the future. 

In this paper, we mainly focus on resource metrics. However, there are many other monitoring data for an application. For example, DApps have blockchain-level data like transaction numbers in each peer node or committed blocks, which can be used for service anomaly detection and root cause localization. In addition, to ensure the running of a cloud application, recovery from performance anomalies is also essential. Based on diagnosis results, an automatic response such as scaling VMs or migrating services to solve anomalies in real-time before users realize it is needed. We can see that an automatic operation system, including monitoring, diagnosis, and recovery, is needed for a cloud application. This paper provides a novel performance diagnosis framework for the operation system. More work needs to be done for cloud application performance management in the future.

\section{Conclusion}
\label{con}
In this paper, we present an integrated performance diagnosis framework named FIRED, which can effectively detect performance anomalies and localize root causes of cloud applications. The performance anomaly detection achieves a better detection accuracy, robustness, and multi-step prediction capability by the deep ensemble method. The proposed root cause localization method can identify root causes in a metric granularity with high accuracy. The FIRED framework focuses on weakly-supervised learning, considering fewer labels exist in real scenarios and provides metrics selection to reduce data dimensions and improve diagnosis performance. We provide experiments to evaluate the effect of metric selection, and results show that it can help improve detection accuracy and reduce time spent. 

We propose the deep ensemble method for performance detection with the requirements of detection accuracy, algorithm robustness, and multi-step prediction. Based on our survey, many detection methods have been developed, but they have different performances because they focus on different data features. Therefore, we propose a deep ensemble method that integrates existing detection methods non-linearly based on ensemble learning. Our experiments compare the deep ensemble method with other detection methods, and the results show that it has the highest detection accuracy and the best algorithm robustness. We also evaluate its multi-step prediction ability, and we can see that it can predict anomalies in four minutes with high accuracy. Finally, we inspect the impact of amounts of labels in the deep ensemble method, and the results show that the method works very well with only fewer labels. Our experiments prove that the deep ensemble method has the highest detection accuracy, best algorithm robustness, and can predict anomalies in four minutes with F1 score higher than 0.8. 

We provide the root cause localization pipeline to fine-grained identify the root causes of performance anomalies accurately and in real-time. The pipeline includes building the dependency graph with the PC algorithm, localizing and ranking root causes with a random walk. We apply the localization pipeline to the DApp monitoring data. We compare the PC and ANM algorithms to build the dependency graph, and the results show that the PC algorithm has the average localization accuracy higher than 0.7. Our experiments also demonstrate the feasibility of real-time root cause localization based on the PC algorithm. More research into improving localization accuracy can be considered in the future.

In future, we will improve detection accuracy and efficiency for the deep ensemble method. As for root cause localization, graph methods which can extract non-linear relations between monitoring data will be developed next. In addition, the diagnosis results based on FIRED framework can be used for exploiting operation strategies to implement rapid recovery. Therefore, an automatic operation system for cloud applications which includes monitoring, diagnosis, and adaptation strategies can be developed in the future. 


\ifCLASSOPTIONcompsoc
  \section*{Acknowledgments}
\else
  \section*{Acknowledgment}
\fi

This research is funded by the EU Horizon 2020 research and innovation program under grant agreements 825134 (ARTICONF project), 862409 (BlueCloud project), 824068 (ENVRIFAIR project), and China Scholarship Council. The research is also funded by Science and Technology Program of Sichuan Province under Grant 2020JDRC0067 and 2020YFG0326. 

\ifCLASSOPTIONcaptionsoff
  \newpage
\fi

\bibliographystyle{IEEEtran}
\bibliography{paper.bib}
%

%

\begin{IEEEbiography}[{\includegraphics[width=1in,height=1.25in,clip,keepaspectratio]{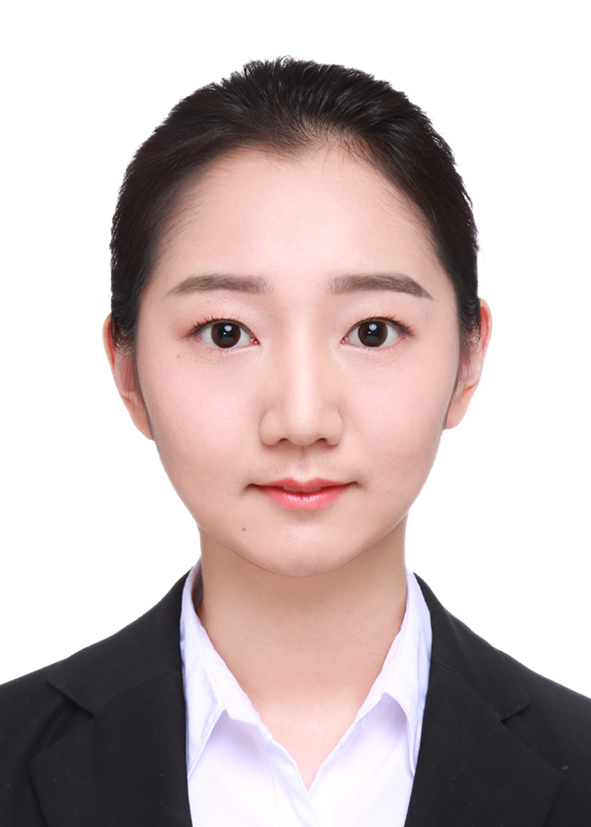}}]{Ruyue Xin}
is currently a Ph.D. candidate in the Multiscale Networked Systems (MNS) research group, University of Amsterdam, the Netherlands. She received the M.Sc. degree in the School of Systems Science, Beijing Normal University, Beijing, China in 2019. Her research interests include AIOps, cloud computing, causality inference.  
\end{IEEEbiography}

\begin{IEEEbiography}[{\includegraphics[width=1in,height=1.25in,clip,keepaspectratio]{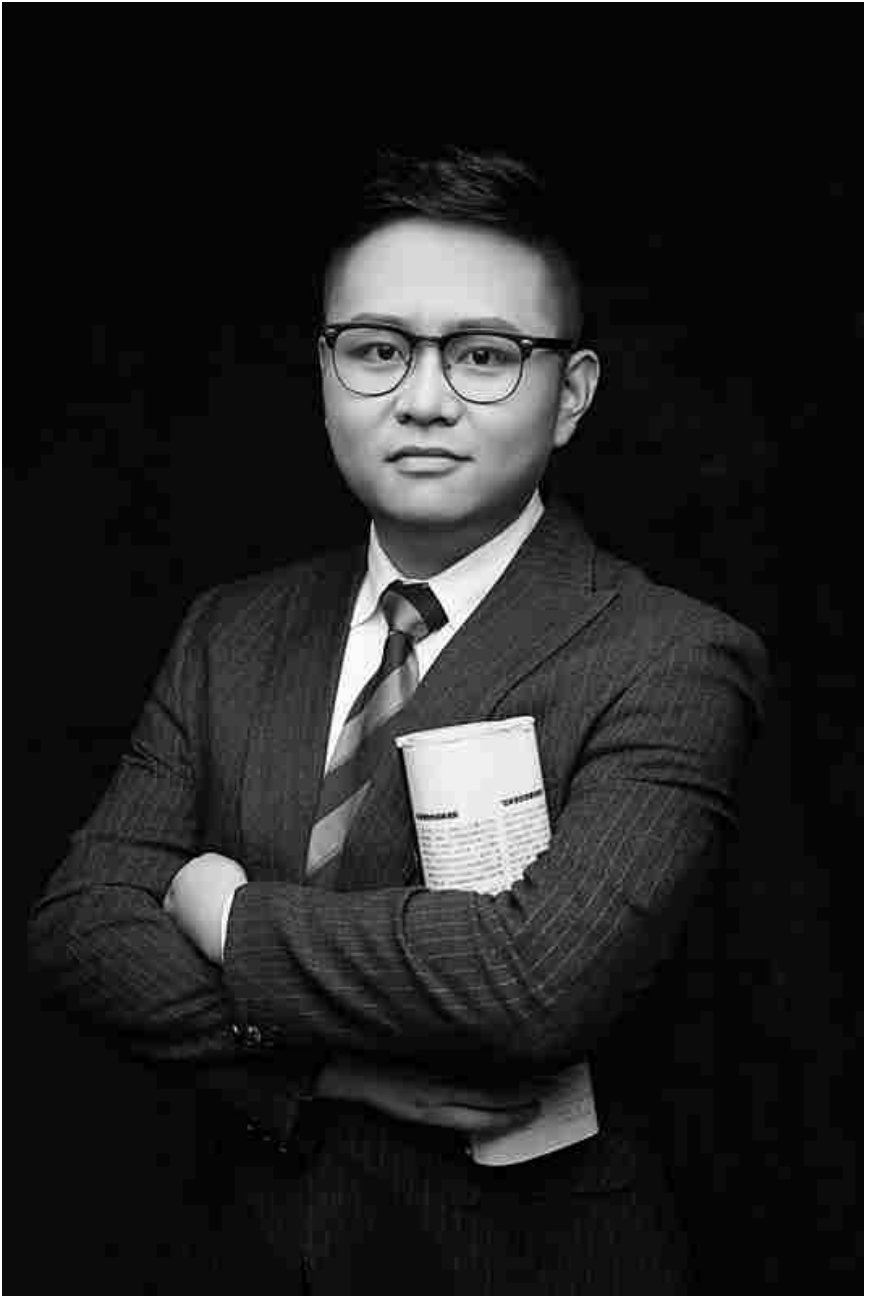}}]{Hongyun Liu}
is currently a Ph.D. candidate in the Multiscale Networked Systems(MNS) research group. He received the BSc degree in Automation and MSc in Navigation, Guidance, and Control both from Northwestern Polytechnical University, Xi'an, China, in 2013 and 2017. His research interests include resource management, cloud computing, applied machine learning.  
\end{IEEEbiography}

\begin{IEEEbiography}[{\includegraphics[width=1in,height=1.25in,clip,keepaspectratio]{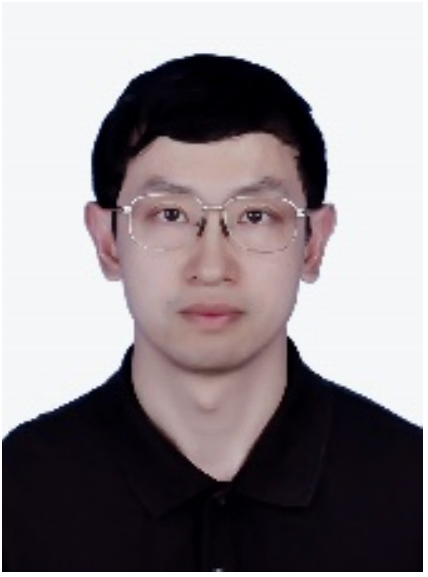}}]{Peng Chen}
is currently a professor with School of Computer and Software Engineering, Xihua University, China. He also was a visiting scholar in the Multiscale Networked Systems (MNS) research group, University of Amsterdam, the Netherlands in 2021. He received the B.E. degree in computer science from University of Electronic Science and Technology of China, Chengdu, China in 2001, the M.Sc. degree in computer software and theory from Peking University, Beijing, China in 2004 and Ph.D. degree in computer science from Sichuan University, Chengdu, China in 2017. His research interests include machine learning and service computing. 
\end{IEEEbiography}


\begin{IEEEbiography}[{\includegraphics[width=1in,height=1.25in,clip,keepaspectratio]{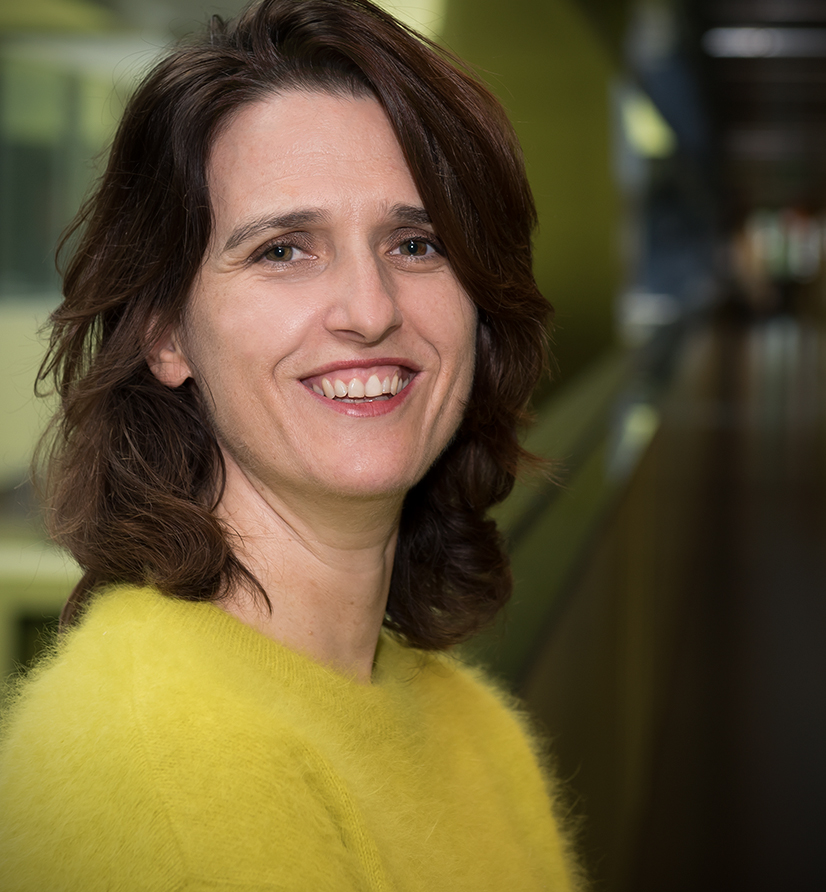}}]{Paola Grosso}
(Member, IEEE) is currently an Associate Professor with the University of Amsterdam where she leads the Multiscale Networked Systems research group (mns-research.nl). Her work focus on the creation of sustainable and secure e-infrastructures, and relying on the provisioning and design of programmable networks. She has an extensive list of publications on the topic and contributes to several national and international projects. 
\end{IEEEbiography}

\begin{IEEEbiography}[{\includegraphics[width=1in,height=1.25in,clip,keepaspectratio]{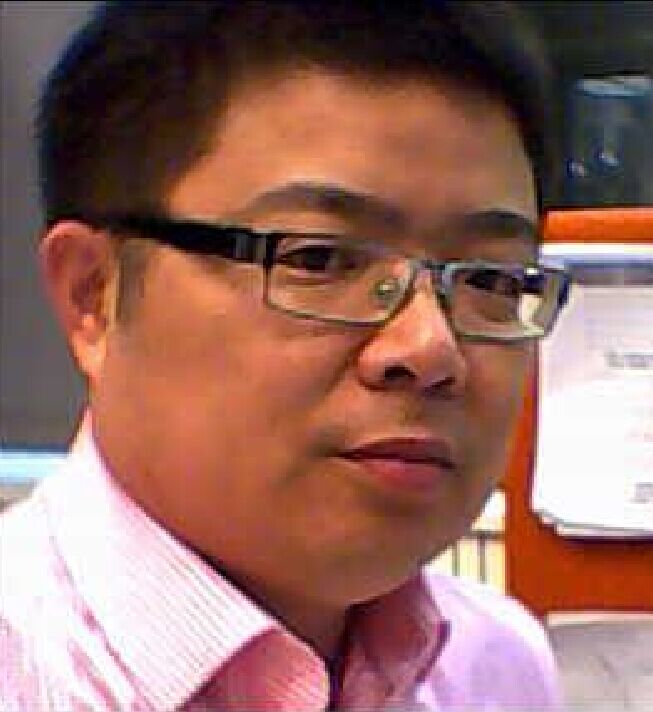}}]{Zhiming Zhao}
is currently an assistant professor in the Multiscale Networked Systems (MNS) research group, University of Amsterdam, the Netherlands. He leads a team on "Quality Critical Distributed Computing" in the System and Networking Lab (SNE). His research focuses on innovative programming and control models. 
\end{IEEEbiography}




\end{document}